\documentclass[12pt]{article}

 \pdfoutput=1

\usepackage{epsfig}

\usepackage{a4wide}
\usepackage{amssymb}
\usepackage{graphicx}
\usepackage{caption}
\usepackage{subcaption}
\usepackage{mathdots}
\usepackage{color}
\usepackage{mathtools}

\def\drawbox#1#2{\hrule height#2pt 
        \hbox{\vrule width#2pt height#1pt \kern#1pt 
              \vrule width#2pt}
              \hrule height#2pt}

\def\Fund#1#2{\vcenter{\vbox{\drawbox{#1}{#2}}}}
\def\Asym#1#2{\vcenter{\vbox{\drawbox{#1}{#2}
              \kern-#2pt       
              \drawbox{#1}{#2}}}}

\def\funda{\Fund{6.5}{0.4}}

\def\symm{\funda\kern-0.4pt\funda}

\def\makeatletter{\catcode`\@=11}
\makeatletter
\def\mathbox#1{\hbox{$\m@th#1$}}%
\def\math@ccstyles#1#2#3#4#5#6#7{{\leavevmode
      \setbox0\mathbox{#6#7}%
      \setbox2\mathbox{#4#5}%
      \dimen@ #3%
      \baselineskip\z@\lineskiplimit#1\lineskip\z@
      \vbox{\ialign{##\crcr
             \hfil \kern #2\box2 \hfil\crcr
             \noalign{\kern\dimen@}%
             \hfil\box0\hfil\crcr}}}}
\def\mathaccstyles{\math@ccstyles\maxdimen}
\def\maththroughstyles{\math@ccstyles{-\maxdimen}}
\def\unity%
 {\maththroughstyles{.45\ht0}\z@\displaystyle {\mathchar"006C}\displaystyle 1}

\def\be{\begin{eqnarray}}
\def\ee{\end{eqnarray}}

\begin{document}

\setcounter{table}{0}

\mbox{}
\vspace{2truecm}
\linespread{1.1}

\begin{center}
{\LARGE \bf Branes and 2d ${\cal N}=(2,2)$ gauge theories with orthogonal and symplectic groups}
\end{center}

\vspace{2truecm}

\centerline{
    {\large \bf Oren Bergman ${}^{a}$} \footnote{bergman@physics.technion.ac.il}
     {\bf and}
    {\large \bf Eran Avraham${}^{a}$} \footnote{eranav@post.bgu.ac.il}}

\vspace{1cm}
\centerline{{\it ${}^a$ Department of Physics, Technion, Israel Institute of Technology}} \centerline{{\it Haifa, 32000, Israel}}
\vspace{1cm}

\centerline{\bf ABSTRACT}
\vspace{1truecm}

\noindent 
We construct two-dimensional ${\cal N}=(2,2)$ supersymmetric gauge theories with orthogonal and symplectic groups
using branes and orientifold planes in Type IIA string theory.
A number of puzzles regarding the construction, including the effect of exchanging NS5-branes
on an orientifold 2-plane, are resolved by lifting the configurations to M theory.
The low energy properties and conjectured dualities of these theories are reproduced in the M-brane description.
A similar construction of ${\cal N}=(4,4)$ theories with orthogonal and symplectic groups 
leads to new duality conjectures for these theories.

\newpage

\tableofcontents

\section{Introduction}

The interplay of gauge theory dynamics and branes in string theory has led to many important developments in 
both quantum field theory and string theory.
In this paper we will fill a long-standing gap in the subject related to two-dimensional ${\cal N}=(2,2)$ supersymmetric gauge theories.

The study of two dimensional ${\cal N}=(2,2)$ gauge theories was motivated mainly by
their role in superstring compactifications \cite{Witten:1993yc}, although they also play a prominent role 
in the description of self-dual strings in six-dimensional superconformal field theories.
Abelian theories are well understood and flow to sigma models whose target space describes toric geometry. 
The dynamics of non-abelian theories are more complicated since they become strongly interacting at low energy.
Non-abelian ${\cal N}=(2,2)$ gauge theories are expected to flow in the IR to sigma models whose target spaces are more general Calabi-Yau manifolds. 
Unitary gauge theories  
were studied in \cite{Hanany:1997vm} and \cite{Hori:2006dk}. 
Orthogonal and symplectic gauge theories were subsequently considered in \cite{Hori:2011pd}.
All these theories exhibit phenomena familiar in four and three dimensions, such as supersymmetry breaking and Seiberg duality.  

A brane construction for ${\cal N}=(2,2)$ theories with a $U(k)$ gauge symmetry and matter in the fundamental representation
was given in \cite{Hanany:1997vm}.
This is a two-dimensional version of the construction of \cite{Elitzur:1997fh} for four-dimensional ${\cal N}=1$
gauge theories (building on the original construction of three-dimensional ${\cal N}=4$ theories in \cite{Hanany:1996ie}),  
in which the two-dimensional gauge theory is realized as the low-energy effective theory on 
D2-branes that are suspended between a pair of NS5-branes, in the presence of ``flavor" D4-branes.
As in the four-dimensional case, the brane construction of the two-dimensional theories is used to exhibit their IR properties.
In particular supersymmetry breaking occurs when there are too many D2-branes suspended between an NS5-brane
and a D4-brane, violating the so-called ``s-rule", and Seiberg duality is realized by exchanging the positions of the two NS5-branes.

Somewhat surprisingly, this construction has not been generalized to orthogonal and symplectic gauge theories in two dimensions.
Brane constructions for four dimensional ${\cal N}=1$ gauge theories with $O(k)$ and $USp(2k)$ groups 
were given in \cite{Evans:1997hk,Elitzur:1997hc}.
But the analogous constructions for two dimensional ${\cal N}=(2,2)$ theories with $O(k)$ and $USp(2k)$ groups 
have not been studied.
Our main goal here is to fill this gap.

The plan of the paper is as follows.
In section 2 we very briefly review the general properties of two-dimensional ${\cal N}=(2,2)$ supersymmetric gauge theories. 
In section 3 we review the brane construction of the $U(k)$ theories.
Section 4 contains our new results related to the brane construction of $O(k)$ and $USp(k)$ theories.
We begin in section 4 by reviewing the field theory results of \cite{Hori:2011pd}, and then present the brane construction 
in Type IIA string theory, and its lift to M theory.
The latter is important in order to resolve a number of puzzles related to
the string theory brane construction.
It will also lead to new insights about brane dynamics in M theory and its reduction to Type IIA string theory.
In section 5 we discuss the generalization to ${\cal N}=(4,4)$ supersymmetric theories.
Section 6 contains our conclusions and open questions.

\section{Basics of 2d ${\cal N}=(2,2)$ supersymmetry}

We begin with a brief review of the basic ingredients of two-dimensional ${\cal N}=(2,2)$
gauge theories 
following \cite{Witten:1993yc,Hanany:1997vm,Hori:2006dk}.
${\cal N}=(2,2)$ supersymmetry has 
supercharges $Q_+, \bar{Q}_+$ and $Q_-, \bar{Q}_-$ carrying charges $(\pm 1,0)$ and $(0,\pm 1)$,
respectively, under the R-symmetry $U(1)_R\times U(1)_L$.
The basic superfields are the chiral superfield,
\be
\Phi(x,\theta) = \phi(x) + \theta^+ \psi_+ + \theta^- \psi_- + \theta^+ \theta^- F + \cdots \,,
\ee
and the gauge superfield, given in Wess-Zumino gauge by
\be
V(x,\theta) &=& \theta^- \bar{\theta}^- (A_0 + A_1) +  \theta^+ \bar{\theta}^+ (A_0 - A_1)
- \theta^+\bar{\theta}^- \sigma - \theta^-\bar{\theta}^+ \sigma^\dagger \nonumber \\
&&
-i \theta^2 \bar{\theta}^\alpha \bar{\lambda}_\alpha
+i \bar{\theta}^2 \theta^\alpha \lambda_\alpha
- \frac{1}{2} \theta^2 \bar{\theta}^2 D \,.
\ee
The gauge superfield can be repackaged as a gauge-covariant {\em twisted chiral superfield}:
\be
\Sigma(x,\theta) = \sigma + \theta^+ \bar{\lambda}_+  + \bar{\theta}^- \lambda_+
+ \theta^+ \bar{\theta}^- (D - iF_{01}) + \cdots \,,
\ee
where $\sigma$ is a complex scalar field parameterizing the Coulomb branch.

The generic ${\cal N}=(2,2)$ gauge theory is given by
\be
{\cal L} = 
 \int d^4\theta \left[ \mbox{Tr}(\bar{\Sigma} \Sigma)
+ \bar{\Phi}_i e^{V_{(i)}} \Phi_i \right]
+ \int d\theta^+ d\theta^- W(\Phi_i) + \mbox{h.c.}
+ \int d\theta^+ d\bar{\theta}^- \, \widetilde{W}(\Sigma) + \mbox{h.c.}
\ee
where the last term is known as a {\em twisted superpotential} interaction.
$W$ has R-charges $(1,1)$ and $\widetilde{W}$ has R-charges $(-1,1)$.
To preserve $U(1)_A = U(1)_{L-R}$, at least at the classical level, the twisted superpotential must be linear in $\Sigma$:
\be 
\label{FItwistedsuperpotential}
\widetilde{W}_{class} = t\, \mbox{Tr}\, \Sigma \,,
\ee
where $t = \xi + i\theta$. The real part is the FI term and the imaginary part is the 2d theta parameter.
In particular $\theta \sim \theta + 2\pi$.
This term exists if the gauge group contains a $U(1)$ factor.
Unlike the superpotential, the twisted superpotential is corrected perturbatively at one loop.
The effective twisted superpotential on the Coulomb branch is given by
\be 
\widetilde{W}_{eff} = \widetilde{W}_{class} + 
 \sum_{i,\vec{w}_i} w_i^\alpha \Sigma^\alpha (\ln(w_i^\alpha\Sigma^\alpha) - 1)
+ \sum_{\vec{r}} r^\alpha \Sigma^\alpha (\ln(r^\alpha\Sigma^\alpha) - 1) 
\ee
where $\alpha = 1, \ldots , \mbox{rank}(G)$,
$\vec{w}_i$ are the weight vectors of the representation of $\Phi_i$ under $G$,
and $\vec{r}$ are the root vectors of $G$.
As a consequence the Coulomb branch is generically lifted.

There are two types of mass terms one can turn on for the matter fields. 
The first is the {\em complex mass} given by the superpotential term
\be 
{\cal L}_m = m \int d\theta^+ d\theta^- \, \tilde{\Phi} \Phi + \mbox{h.c.}
\ee
This is just the reduction of the 4d ${\cal N}=1$ mass term.
The second is the {\em twisted mass} (which is also complex) given by 
\be 
{\cal L}_{\tilde{m}} = \int d^4\theta \, \Phi^\dagger \, e^{\theta^+\bar{\theta}^-\tilde{m}+{h.c.}} \, \Phi \,.
\ee
This can be thought of as a VEV for a scalar in a background vector multiplet associated to the global flavor symmetry.
The twisted mass in two dimensions is also related to the real mass in three dimensions.

\section{Review $U(k)$ theories}
\subsection{Field theory}

Two dimensional ${\cal N}=(2,2)$ gauge theories with $U(k)$ gauge symmetry 
and matter in the fundamental representation
were studied in \cite{Hanany:1997vm,Hori:2006dk}.
With $n_f$ fundamentals and $n_a$ anti-fundamentals, the global flavor symmetry is $SU(n_f)\times SU(n_a) \times U(1)_-$.
The effective twisted superpotential on the Coulomb branch is given by
\be
\label{EffectiveTwistedW}
\widetilde{W}_{eff} = \left(t + i\pi(k-1)\right)\mbox{Tr}\, \Sigma + n_f \, \mbox{Tr} [\Sigma(\ln\Sigma - 1)]
- n_a \, \mbox{Tr} [\Sigma(\ln(-\Sigma) - 1)] \,.
\ee
Note that in two dimensions the chiral anomaly does not require $n_f=n_a$.
For $n_f \neq n_a$ the theories are massive, and the $U(1)_A$ R-symmetry is broken anomalously to $\mathbb{Z}_{2(n_f-n_a)}$.
We will be mostly interested in the case $n_f = n_a = n$.
In this case the $U(1)_A$ R-symmetry is unbroken, and the theory has a non-trivial IR fixed point.
The effective twisted superpotential reduces to
\be
\widetilde{W}_{eff} = \left(t + i\pi(n+k-1)\right)\mbox{Tr}\, \Sigma \,,
\ee
which amounts to just a shift in the theta parameter $\theta_{eff} = \theta + \pi(n+k-1)$.

For $n\leq k$ the theory has no supersymmetric ground states.
For $n\geq k+1$ supersymmetry is unbroken and there are $n\choose k$ supersymmetric ground states.\footnote{For $SU(k)$
this is slightly different \cite{Hori:2006dk}, but we will mostly be concerned with $U(k)$.}
For $n=k+1$ the theory flows in the IR to a free theory of baryons and mesons.
For $n>k+1$ there is a dual ``magnetic" theory with a gauge group $U(n-k)$, $n$ fundamentals $q$, $n$ anti-fundamentals $\tilde{q}$,
and a singlet $M$ in the bi-fundamental representation of the flavor symmetry, with a superpotential 
$W = \tilde{q} M q$. In fact this is a special case of a more general duality for $n_f$ fundamentals and 
$n_a$ antifundamentals \cite{Benini:2012ui}.
The duality for $n_a = 0$ was originally proposed in \cite{Hanany:1997vm}.
Evidence for these dualities was provided by comparing the $S^2$ partition functions in \cite{Benini:2012ui,Doroud:2012xw,Benini:2014mia,Gomis:2014eya},
and by comparing the elliptic genera in \cite{Benini:2013xpa}.

\subsection{Branes}

\begin{figure}[h!]
\center
\includegraphics[height=0.3\textwidth]{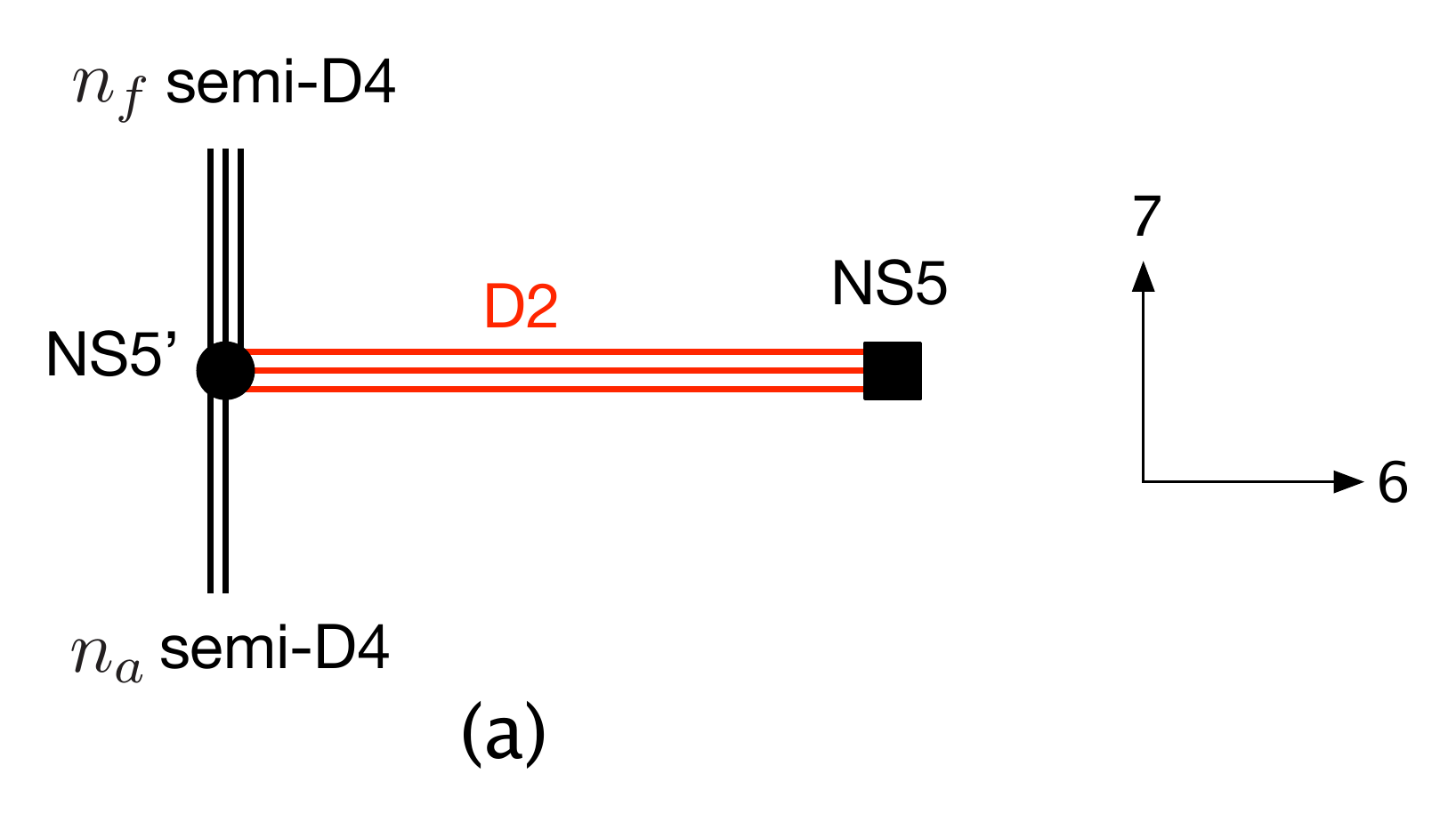} 
\hspace{10pt}
\includegraphics[height=0.3\textwidth]{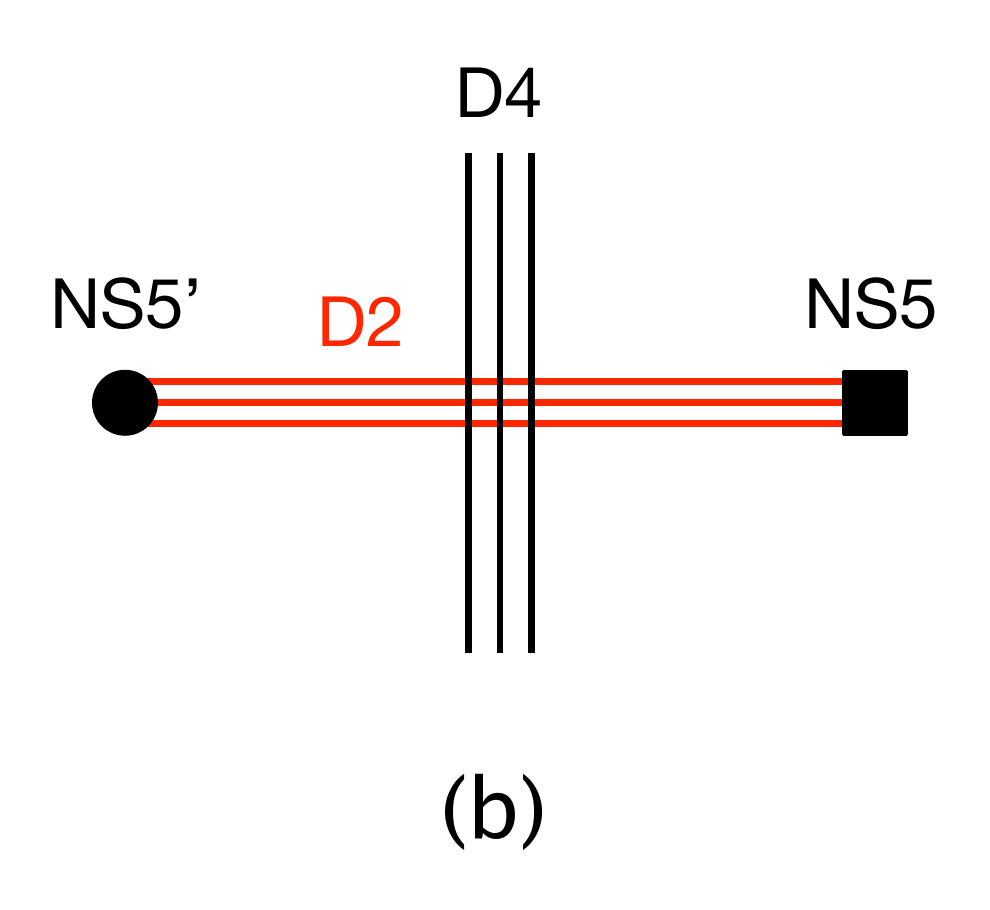} 
\caption{(a) Type IIA brane configuration for $U(k)$ with $n_f$ fundamentals and $n_a$ anti-fundamentals.
(b) $n_f=n_a$ case.}
\label{IIAbranes1}
\end{figure}

The brane construction of \cite{Hanany:1997vm} for the 2d ${\cal N}=(2,2)$ theories with $G=U(k)$ is summarized in Table~\ref{IIAbranescan} 
and shown in Fig.~\ref{IIAbranes1}.
In particular, each semi-infinite D4-brane ending on the NS5'-brane from $x^7 > 0$ provides a chiral superfield in the fundamental representation,
and each semi-infinite D4-brane ending on the NS5'-brane from $x^7 < 0$ provides a chiral superfield in the anti-fundamental representation.
From the D2-brane point of view there is a global chiral symmetry $SU(n_f)\times SU(n_a)$, as well as three $U(1)$'s corresponding to rotations
in the (2,3), (4,5) and (8,9) planes. The first two are the $U(1)_A$ and $U(1)_V$ R-symmetries, respectively, and the third is 
the axial $U(1)_-$ part of $U(n_f)\times U(n_a)$ D4-brane gauge symmetry. 
The vector $U(1)_+$ part is contained in the 2d gauge symmetry.
Pairs of semi-infinite D4-branes can connect and move into the interval, breaking $SU(n)\times SU(n)$ to a single $SU(n)$.
From the D4-brane point of view this corresponds to giving a VEV to a bi-fundamental field.

\begin{table}[h!]
\begin{center}
\begin{tabular}{|l|l|l|l|l|l|l|l|l|l|l|}
 \hline 
 brane& 0 & 1 & 2 & 3 & 4 & 5 & 6 & 7 & 8 & 9 \\
 \hline 
 NS5 & $\bullet$  & $\bullet$& $\bullet$& $\bullet$& $\bullet$& $\bullet$  & & & &   \\
 NS5' & $\bullet$  & $\bullet$& $\bullet$& $\bullet$& & & & & $\bullet$ & $\bullet$  \\
 D2 & $\bullet$ & $\bullet$ & & & & & $\bullet$ & & & \\
 D4 & $\bullet$ & $\bullet$ & & & & & & $\bullet$ & $\bullet$ & $\bullet$ \\
 \hline
 \end{tabular}
 \end{center}
\caption{Type IIA brane configuration for $U(k)$}
\label{IIAbranescan}
\end{table}

\begin{figure}[h!]
\center
\includegraphics[height=0.3\textwidth]{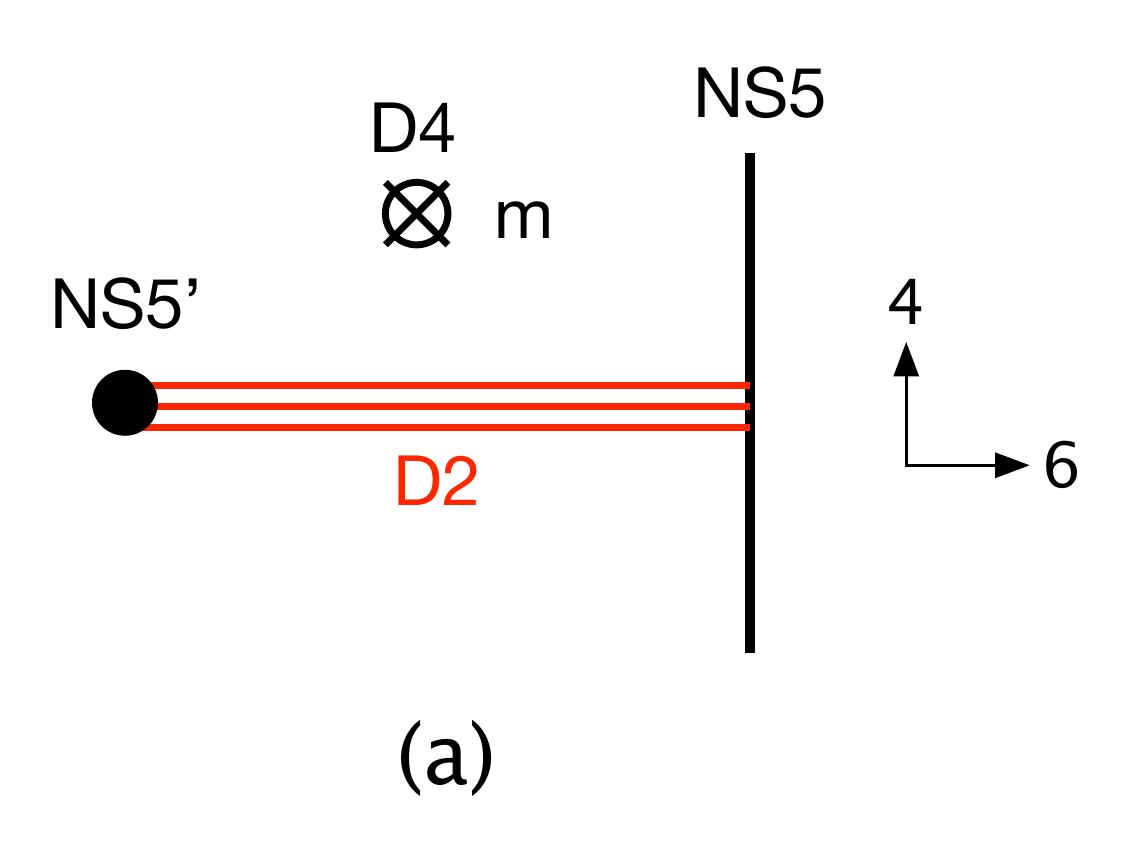} 
\hspace{30pt}
\includegraphics[height=0.3\textwidth]{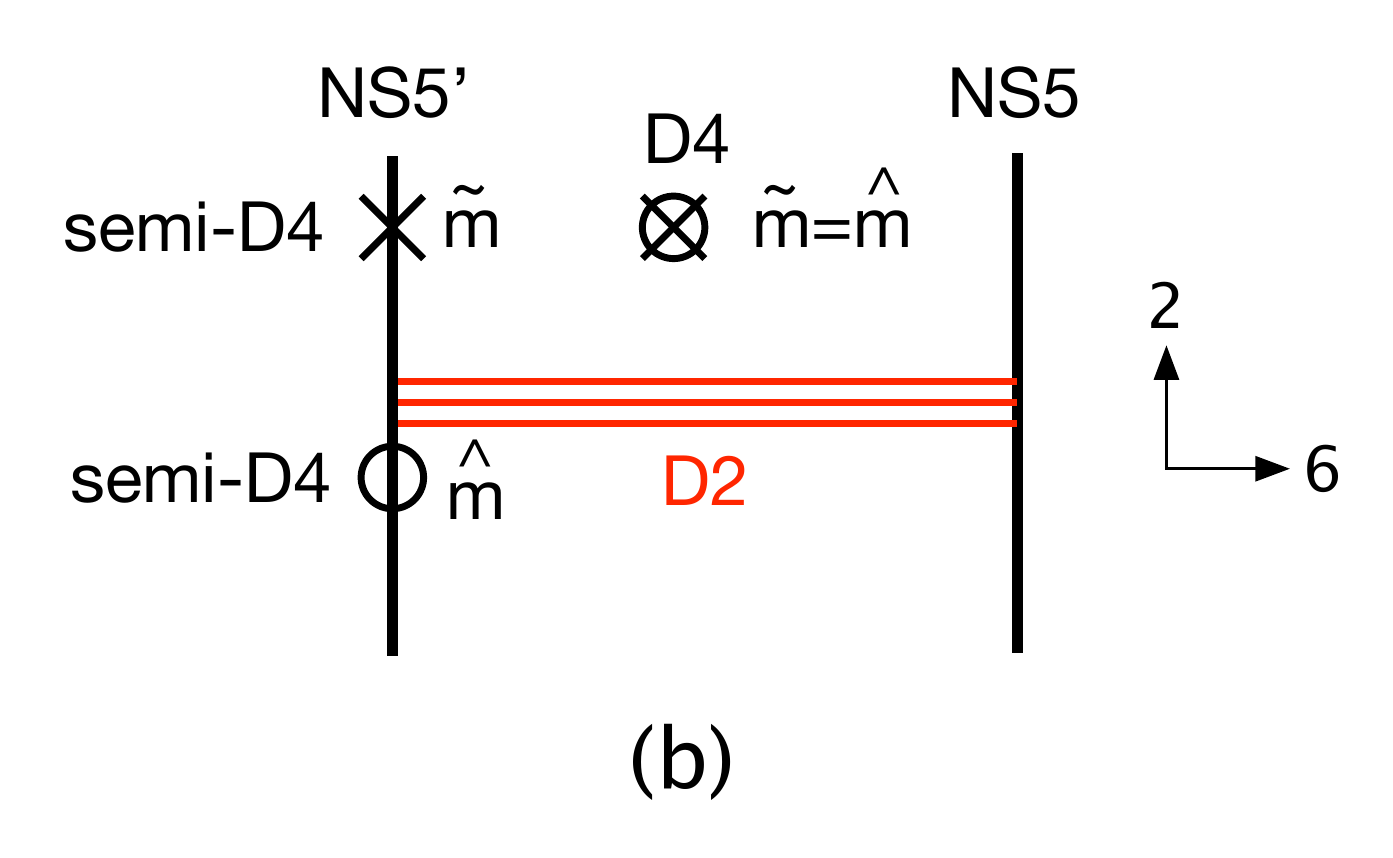} 
\caption{Mass deformations: (a) Complex mass (b) Twisted mass}
\label{IIAbranes2}
\end{figure}

\begin{figure}[h!]
\center
\includegraphics[height=0.3\textwidth]{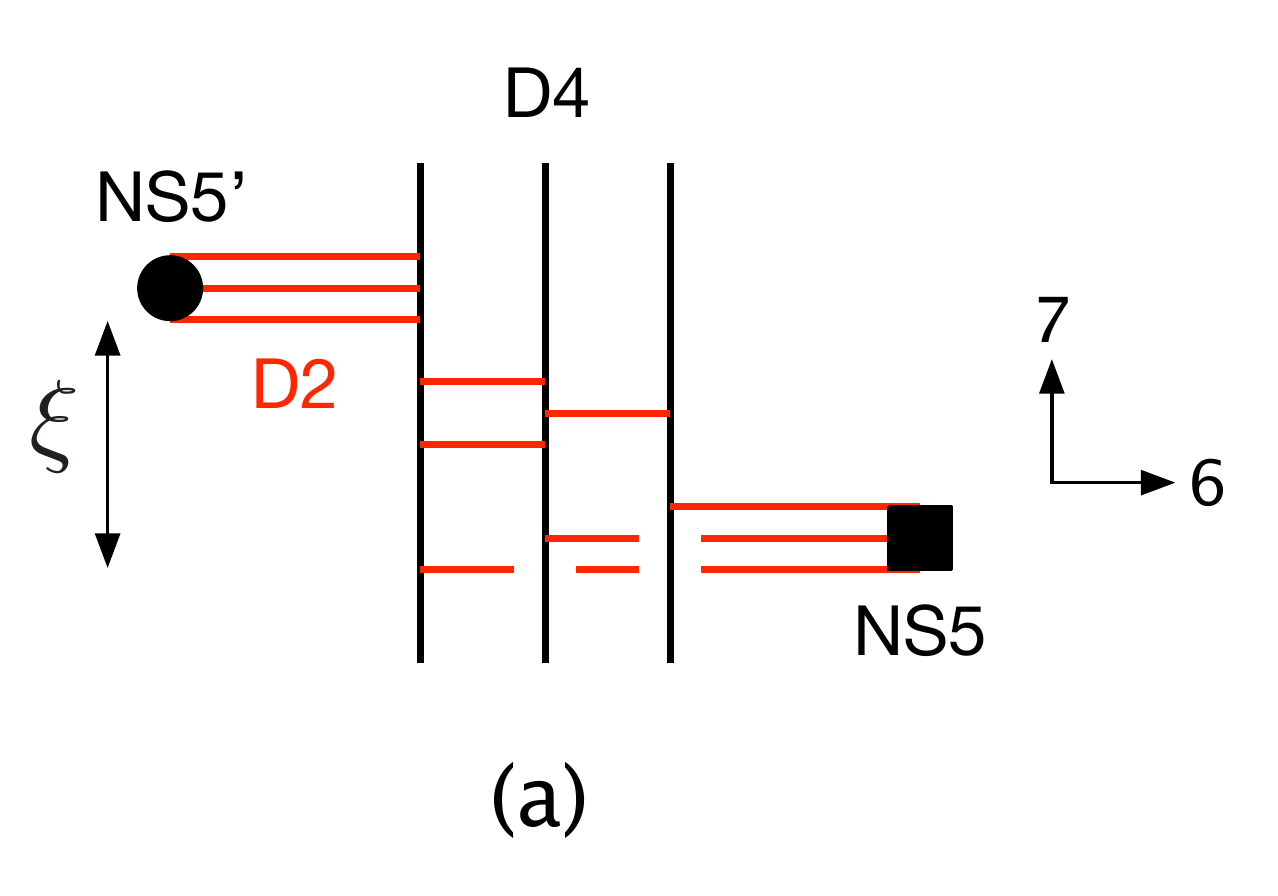} 
\hspace{30pt}
\includegraphics[height=0.32\textwidth]{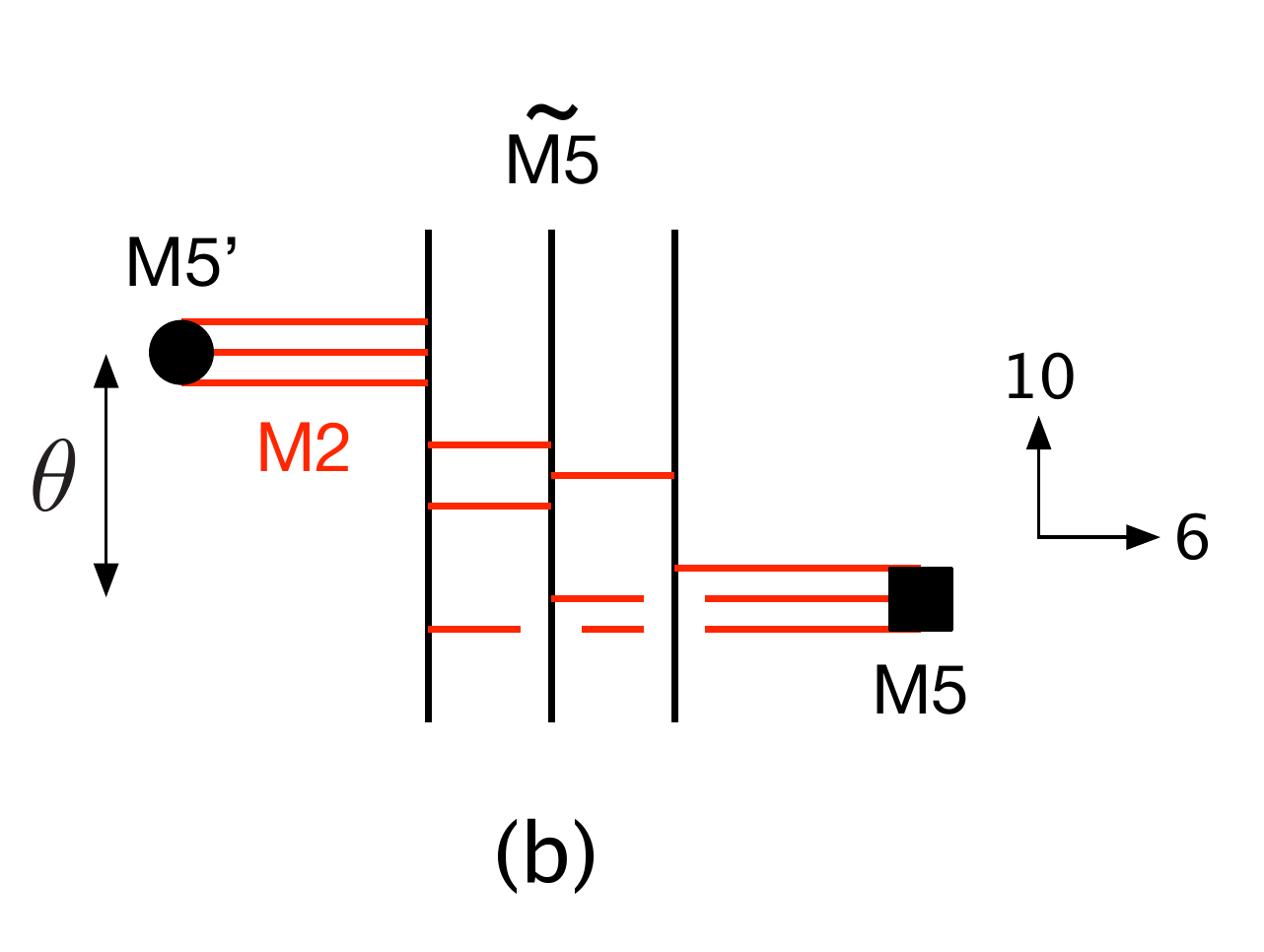} 
\caption{FI and $\theta$ parameters}
\label{IIAbranes3}
\end{figure}

Let us focus on the configuration with $n_f=n_a = n$. 
In this case all the D4-branes can be connected and moved into the interval, Fig.~\ref{IIAbranes1}b.
The moduli space and parameters of the gauge theory can be read off from the geometry of the brane configuration.
The positions of the D2-branes in the (2,3) plane correspond to the Coulomb branch of the theory.
The Higgs branch is described by breaking the D2-branes into segments between the D4-branes and moving them in the (7,8,9) directions.
The positions of the D4-branes in the (4,5) plane are complex masses $m_i^{\tilde{j}}$, Fig.~\ref{IIAbranes2}a.
Generic twisted masses $\widetilde{m}_i$, $\widehat{m}_{\tilde{j}}$ are described by breaking D4-branes on the NS5'-brane,
and independently moving the two pieces in the (2,3) plane, Fig~\ref{IIAbranes2}b.
The position of a full D4-brane in the (2,3) plane corresponds to $\widetilde{m}_i = \widehat{m}_{\tilde{i}}$.
The separation of the NS5-branes in $x^6$ is the Yang-Mills coupling $g^{-2}_{YM}$, and their separation in $x^7$
is the FI parameter $\xi$, Fig.~\ref{IIAbranes3}a.
The $\theta$ parameter becomes visible in the M theory lift of the configuration (see Table~\ref{Mbranescan}), 
where it corresponds to the separation of 
the M5-brane and M5'-brane in $x^{10}$, Fig.~\ref{IIAbranes3}b.\footnote{From the point of view of the NS5-branes this corresponds to 
the fifth scalar of the 6d $(2,0)$ tensor multiplet.}
There is one small subtlety in the identification of the $\theta$ parameter: the separation of the two M5-branes should be identified with
$\theta + \pi(k-1)$, namely there is a shift of $\pi$ if $k$ is even.
This provides a unified description of the complex parameter $t = \xi + i\theta$ as the separation of the M5-brane 
and M5'-brane in the (7,10) plane.

\begin{table}[h!]
\begin{center}
\begin{tabular}{|l|l|l|l|l|l|l|l|l|l|l|l|}
 \hline 
 brane& 0 & 1 & 2 & 3 & 4 & 5 & 6 & 7 & 8 & 9 & \textcolor{red}{10}\\
 \hline 
 M5 & $\bullet$  & $\bullet$& $\bullet$& $\bullet$& $\bullet$& $\bullet$  & & & &  & \\
 M5' & $\bullet$  & $\bullet$& $\bullet$& $\bullet$& & & & & $\bullet$ & $\bullet$  & \\
 M2 & $\bullet$ & $\bullet$ & & & & & $\bullet$ & & & & \\
 $\widetilde{\mbox{M5}}$ & $\bullet$ & $\bullet$ & & & & & & $\bullet$ & $\bullet$ & $\bullet$ & \textcolor{red}{$\bullet$}\\
 \hline
 \end{tabular}
 \end{center}
\caption{M-brane configuration}
\label{Mbranescan}
\end{table}

For $t\neq 0$ the ground state is described by a configuration where the M2-branes (D2-branes) break on the $\widetilde{\mbox{M5}}$-branes
(D4-branes), corresponding to the Higgs branch of the theory.
This configuration has $k$ M2-branes suspended between the M5-brane and the $\widetilde{\mbox{M5}}$-branes.
Since the M5-brane and $\widetilde{\mbox{M5}}$-brane are a linked pair one must apply the s-rule, which implies that
supersymmetry is unbroken only if $n\geq k$, and there are $n\choose k$ supersymmetric ground states.
This reproduces the field theory result.

\begin{figure}[h!]
\center
\includegraphics[height=0.22\textwidth]{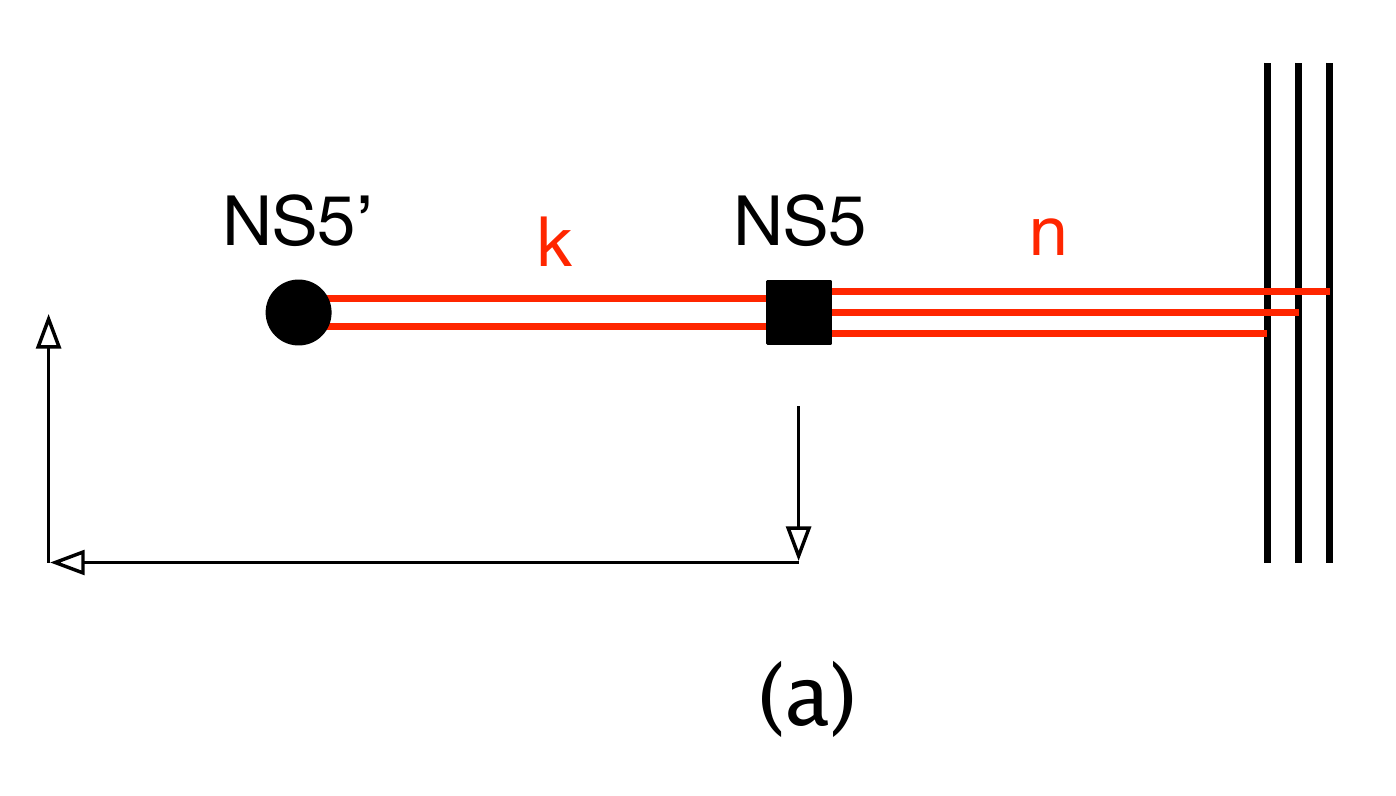} 
\hspace{30pt}
\includegraphics[height=0.22\textwidth]{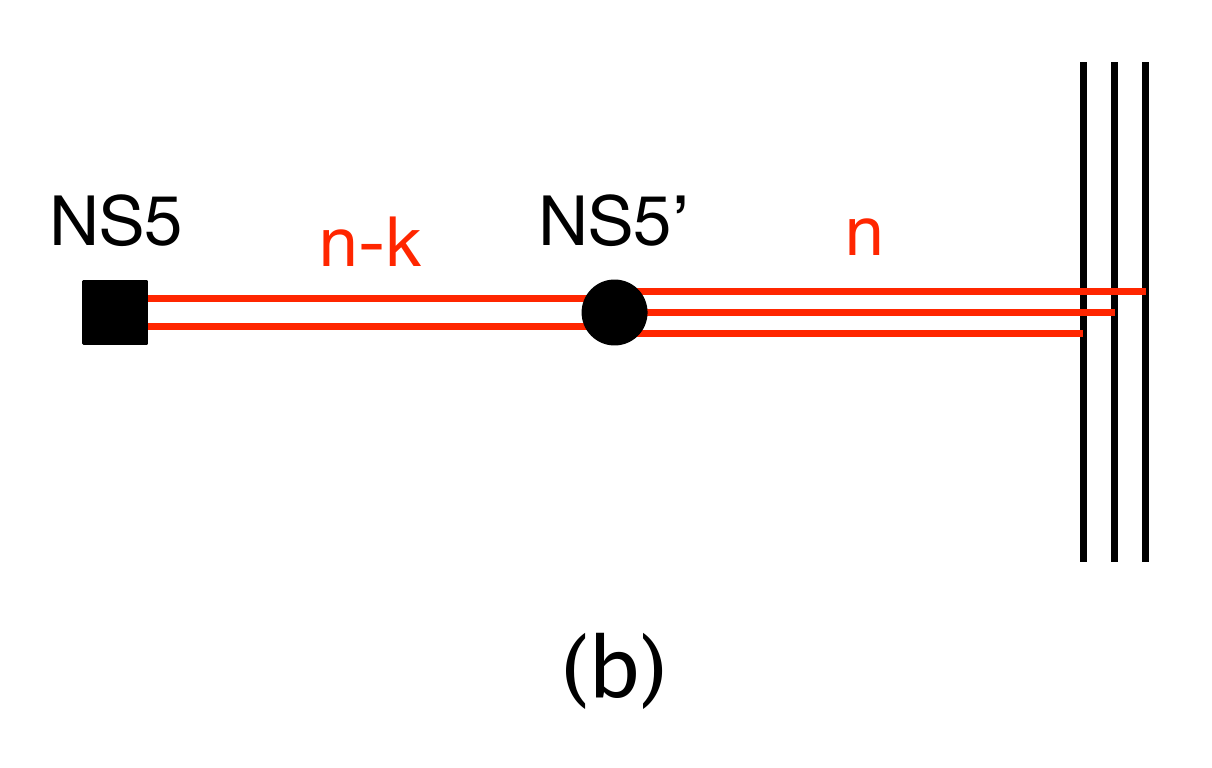}
\caption{The duality move.}
\label{IIAbranes4}
\end{figure}

The dual theory is obtained, as usual, by first moving the D4-branes ($\widetilde{\mbox{M5}}$-branes) to the right and across the 
NS5-brane (M5-brane), leading to the creation of $n$ D2-branes (M2-branes), and then exchanging the NS5-brane and NS5'-brane, keeping $t\neq 0$, Fig.~\ref{IIAbranes4}.
The final configuration, Fig~\ref{IIAbranes4}b, describes the ``magnetic" theory, with $G=U(n-k)$, $n$ fundamentals and anti-fundamentals,
and $n^2$ singlets.

For $n_f\neq n_a$ the story is slightly more involved and one has to take into account the bending of the NS5'-brane,
or more precisely the curve corresponding to the joined M5'-brane and $\widetilde{\mbox{M5}}$-branes \cite{Hanany:1997vm}.
This describes the renormalization of the FI and $\theta$ parameters.
The curve is given by
\be
t(\sigma) = \mbox{const} - \sum_{i=1}^{n_f} \ln(\sigma - \widetilde{m}_i) + \sum_{\tilde{j}=1}^{n_a} \ln(\sigma - \widehat{m}_{\tilde{j}}) \,,
\ee
where $\sigma = x^2 + ix^3$ and $t=x^7 +ix^{10}$,
in agreement with the field theory computation of the effective twisted superpotential (\ref{EffectiveTwistedW}).

\section{Orthogonal and symplectic theories}
\subsection{Field theory}

The 2d ${\cal N}=(2,2)$ theories with orthogonal and symplectic gauge groups were studied by Hori in \cite{Hori:2011pd}.

In the orthogonal case the gauge group is either $O(K)$ or $SO(K)$, with $N$ chiral superfields in the vector representation.
In either case the global flavor symmetry is $U(N)$.
These theories admit a discrete theta parameter $\theta_D\in \{0,\pi\}$ associated to $\pi_1(SO(K)) = \mathbb{Z}_2$.\footnote{This is true for $K\geq 3$.
For $K=2$, $\pi_1(SO(2)) =\mathbb{Z}$.  The $SO(2)=U(1)$ theory admits a real valued theta parameter which contributes
a term $\theta \int F$ to the action.
However since the gauge field is odd under the inversion element of $O(2)$, only $\theta =0, \pi$ are possible in the $O(2)$ theory.}
There is actually an additional discrete choice for the $O(K)$ theory which is understood as follows.
The $O(K)$ theory can be obtained by gauging a $\mathbb{Z}_2$ global symmetry in the $SO(K)$ theory.
This symmetry is either charge conjugation, or charge conjugation combined with $(-1)^F$.
The former is the so-called {\em standard orbifold}, and the latter is the {\em non-standard} orbifold.
The main difference between the two theories is that the number of supersymmetric ground states of the standard
orbifold is doubled relative to the non-standard orbifold.
The notation $O(K)_{\pm}$ was introduced in \cite{Hori:2011pd} to differentiate between the two theories.
However the correspondence depends on $K$ and $N$ as follows: $O(K)_+$ is the standard (non-standard) orbifold for 
$N+K$ odd (even), and $O(K)_-$ is the standard (non-standard) orbifold for $N+K$ even (odd).

The Coulomb branch of the orthogonal theory is parameterized by
\be
\sigma = \left\{
\begin{array}{ll}
\mbox{diag}(\sigma_1 \tau^2, \ldots, \sigma_k \tau^2) & \mbox{for} \; K=2k \\
\mbox{diag}(\sigma_1 \tau^2, \ldots, \sigma_k \tau^2,0) & \mbox{for} \; K=2k+1 \,,
\end{array}
\right.
\ee
and the effective twisted superpotential on the Coulomb branch is given by
\be
\widetilde{W}_{eff} = i\left(\theta_D + \pi(N+K)\right)\sum_{a=1}^k \Sigma_a \,.
\ee
Consequently, even though the theory does not admit an FI parameter, the Coulomb branch is lifted if $N+K$ is odd and $\theta_D=0$,
or if $N+K$ is even and $\theta_D = \pi$.
In \cite{Hori:2011pd} these were referred to as the {\em regular theories}.
Supersymmetry is unbroken if $N\geq K-1$.
The number of supersymmetric ground states depends on the theory and on whether $K$ and $N$ are even or odd (see Table (4.20) in \cite{Hori:2011pd}).
For example, the {\em regular} theory with gauge group $O(2k)_\pm$ and $2n$ flavors has $n\choose k$ supersymmetric ground states.
For $N=K-1$ the theories flow to free theories of mesons. 
For $N\geq K$, Hori proposed the following set of dualities for the {\em regular} theories:
\be
SO(K)&\longleftrightarrow & O(N-K+1)_+\\
O(K)_+&\longleftrightarrow & SO(N-K+1)\\
O(K)_-& \longleftrightarrow & O(N-K+1)_- \,,
\ee
where as in four dimensions, the theory on the RHS of the duality contains in addition to the $N$ flavors $q_i$, 
singlets $s_{ij}$ in the symmetric representation of $U(N)$, and a superpotential
$W=\sum_{i,j=1}^N s_{ij}q_iq_j$.

\medskip

In the symplectic theories the gauge group is $USp(2k) = Sp(k)$, and there are $N$ chiral superfields in the fundamental
($2k$-dimensional) representation. Note that unlike four and three dimensions, $N$ may be odd, since there is no global anomaly
for $Sp(k)$ in two dimensions.
The global symmetry is again $U(N)$.
The Coulomb branch is parametrized by 
\be
\sigma = \mbox{diag}(\sigma_1 \tau^3, \ldots, \sigma_k \tau^3) \,.
\ee
As in the orthogonal theories the quantum effects on the Coulomb branch amount to effective $U(1)$ theta parameters $\theta_{eff}=\pi N$.
Therefore the theory is {\em regular} only if $N$ is odd, $N=2n+1$.
Supersymmetry is unbroken provided that $n\geq k$, and there are $n\choose k$ supersymmetric vacua.
For $n=k$ the theory flows in the IR to a free theory of mesons. For $n\geq k+1$ there is a proposed duality
\be
USp(2k)\longleftrightarrow USp(N-2k-1),
\ee
where the theory on the RHS contains singlets $ a_{ij}$ in the antisymmetric representation of $U(N)$,
and a superpotential, $W=\sum_{i,j=1}^N a_{ij}q_iq_j$.

The duality conjectures for the orthogonal and symplectic theories were originally based on 
't Hooft anomaly matching, the number of supersymmetric ground states, and a comparison of the
$(c,c)$ chiral ring of gauge invariant polynomials of the chiral superfields, and the
$(a,c)$ chiral ring of gauge invariant polynomials of the twisted chiral superfield (the vector superfield) \cite{Hori:2011pd}.
These dualities were also tested by comparing the elliptic genus in \cite{Kim:2017zis}, 
as well as correlation functions of Coulomb branch operators in \cite{Closset:2017vvl}.

\medskip

The orthogonal and symplectic dualities in two dimensions appear to be a natural progression of the analogous dualities
in four and three dimensions, in which there is a shift in the rank of the magnetic theory of $\pm 4$ and $\pm 2$, respectively,
relative to the unitary case.
This suggests an interpretation in terms of orientifold planes, since their charge also decreases by a factor of 2 
for each dimensional reduction.
In what follows we will see that this simple observation is essentially correct, but subtle,
and it will require the perspective of M theory.

\subsection{Type IIA branes} 

\begin{figure}[h!]
\center
\includegraphics[height=0.18\textwidth]{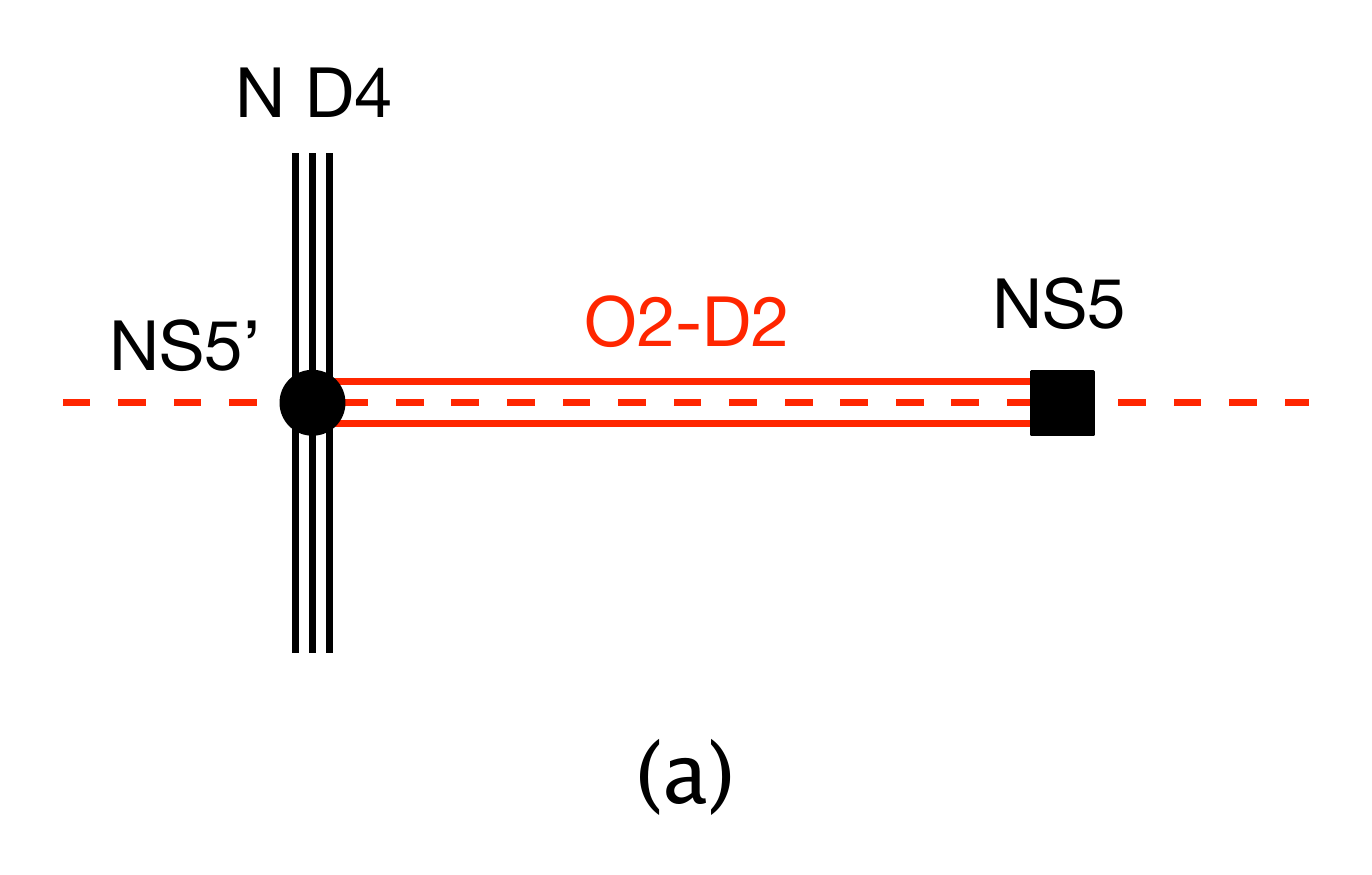} 
\includegraphics[height=0.18\textwidth]{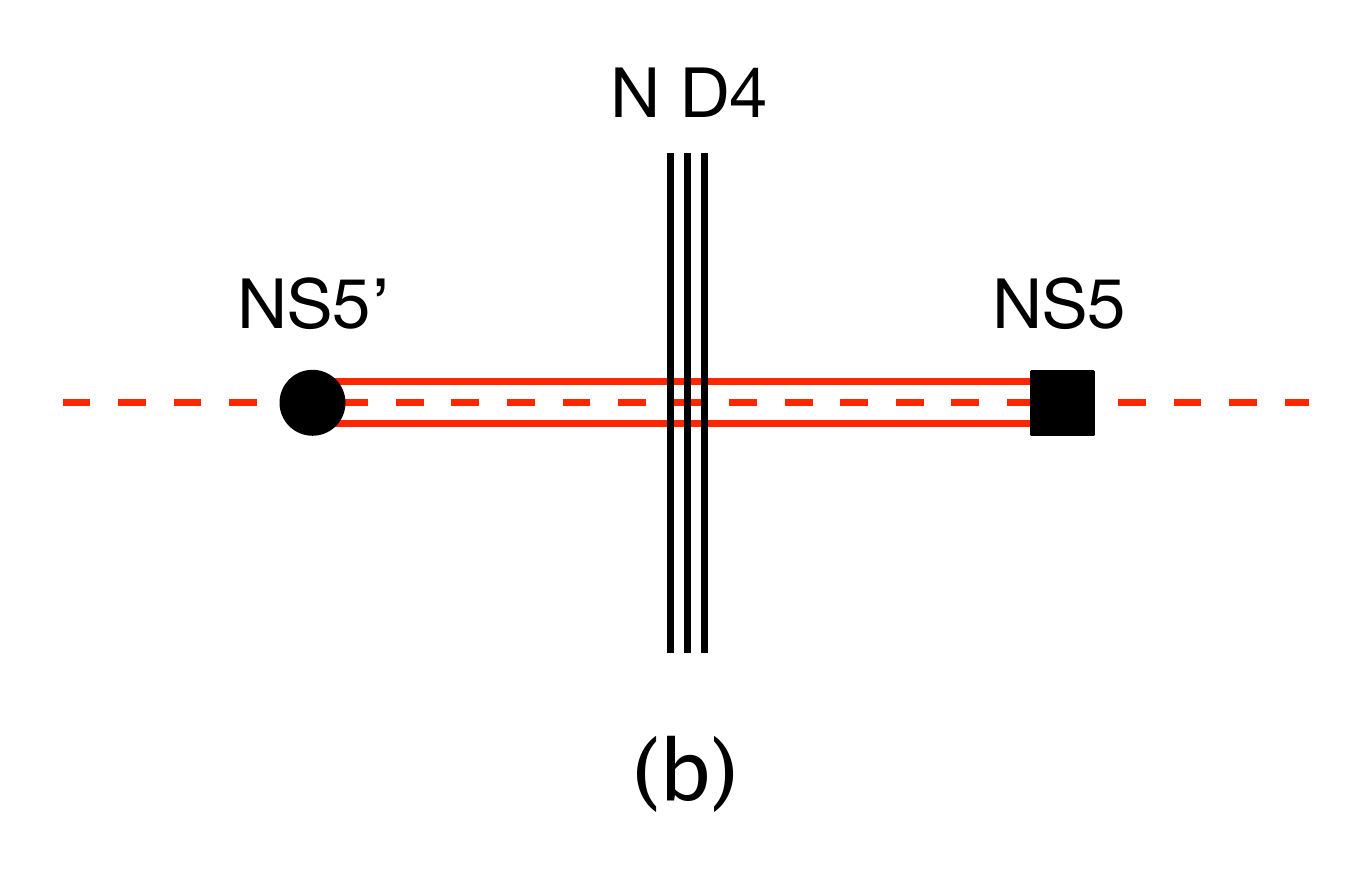}
\includegraphics[height=0.18\textwidth]{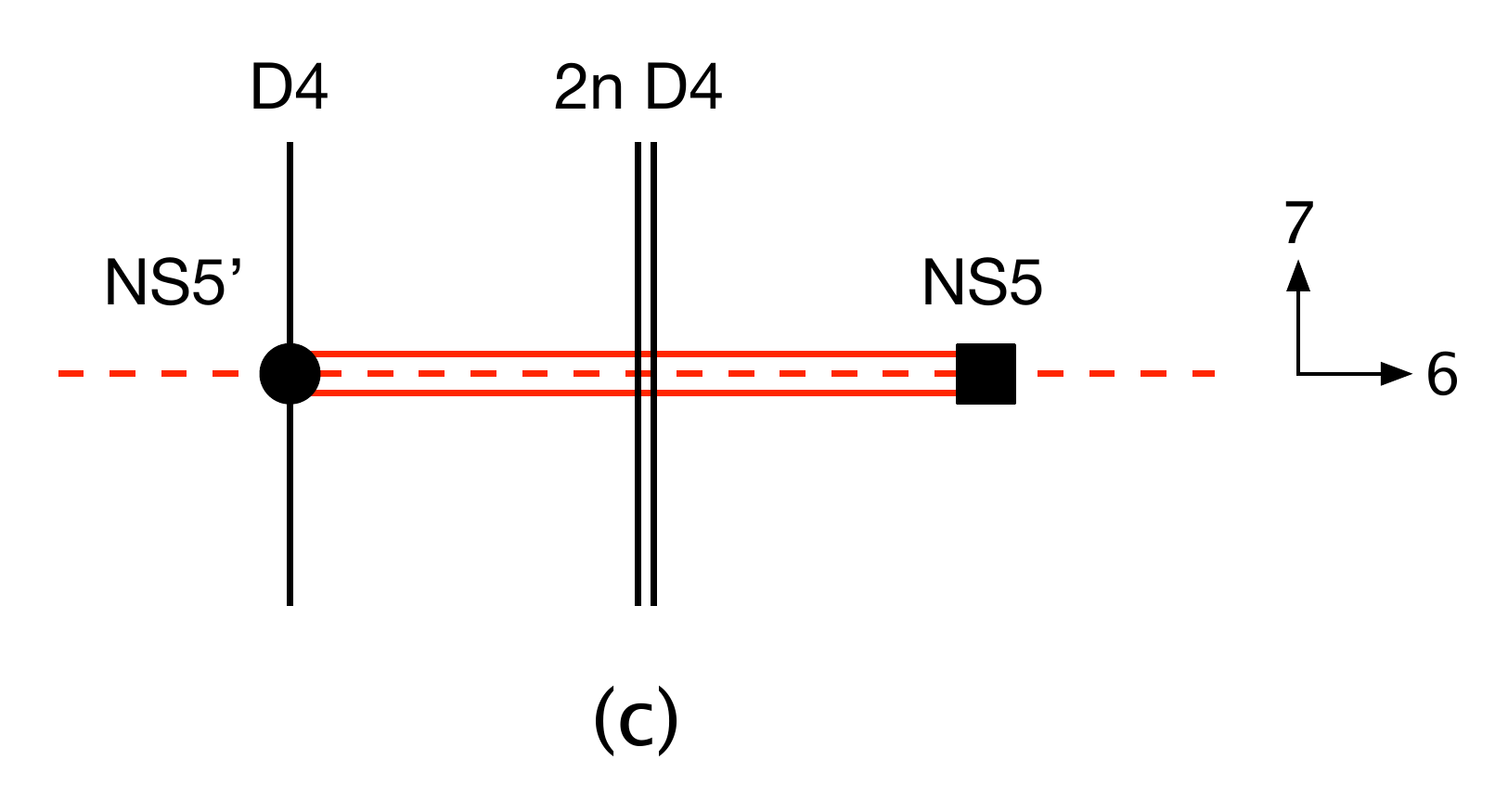}
\caption{Type IIA brane configuration for $O(K)$ and $USp(2k)$. For $O(K)$ with an odd number of flavors one D4-brane 
must remain broken on the NS5'-brane.}
\label{Orientifold1}
\end{figure}

To extend the Type IIA brane construction to the orthogonal and symplectic theories we add an orientifold 2-plane,
as shown in Fig.~\ref{Orientifold1}.
The two NS5-branes are now ``fractional" in the sense that they are fixed to the location of the O2-plane.
Furthermore, the type of O2-plane changes across each NS5-brane, from 
$\mbox{O2}^\pm$ to $\mbox{O2}^\mp$, or from $\widetilde{\mbox{O2}}^\pm$ to 
$\widetilde{\mbox{O2}}^\mp$ \cite{Hanany:2000fq}.
The gauge group of the 2d gauge theory depends on the type of O2-plane between the NS5-branes.
For $\mbox{O2}^-$ the gauge group is $O(2k)$, for $\widetilde{\mbox{O2}}^-$ it is $O(2k+1)$, 
and for $\mbox{O2}^+$ and $\widetilde{\mbox{O2}}^+$ it is $USp(2k)$.
However the brane construction does not distinguish between the $O_+$, $O_-$, and $SO$ theories.
The flavor D4-branes must now have both an $x^7<0$ piece and an $x^7>0$ piece that are related by the orientifold projection.
The global symmetry for $N$ fundamental chiral multiplets is therefore $U(N)$. 
As before, moving the D4-branes into the interval (Fig.~\ref{Orientifold1}b) reduces the global symmetry.
If the orientifold plane in the interval is an $\mbox{O2}^+$ or $\widetilde{\mbox{O2}}^+$, $U(N)$ breaks to $O(N)$.
From the D4-brane point of view, this corresponds to a VEV for a field in the symmetric representation of $U(N)$.
If the orientifold plane in the interval is an $\mbox{O2}^-$ or $\widetilde{\mbox{O2}}^-$, $U(N)$ breaks to $USp(N)$ if $N$ is even,
and to $USp(N-1)\times U(1)$ is $N$ is odd.
In the latter case a single D4-brane must remain on the NS5'-brane, Fig.~\ref{Orientifold1}c.
This corresponds to a VEV for a field in the antisymmetric representation of $U(N)$.

\begin{figure}[h!]
\center
\includegraphics[height=0.2\textwidth]{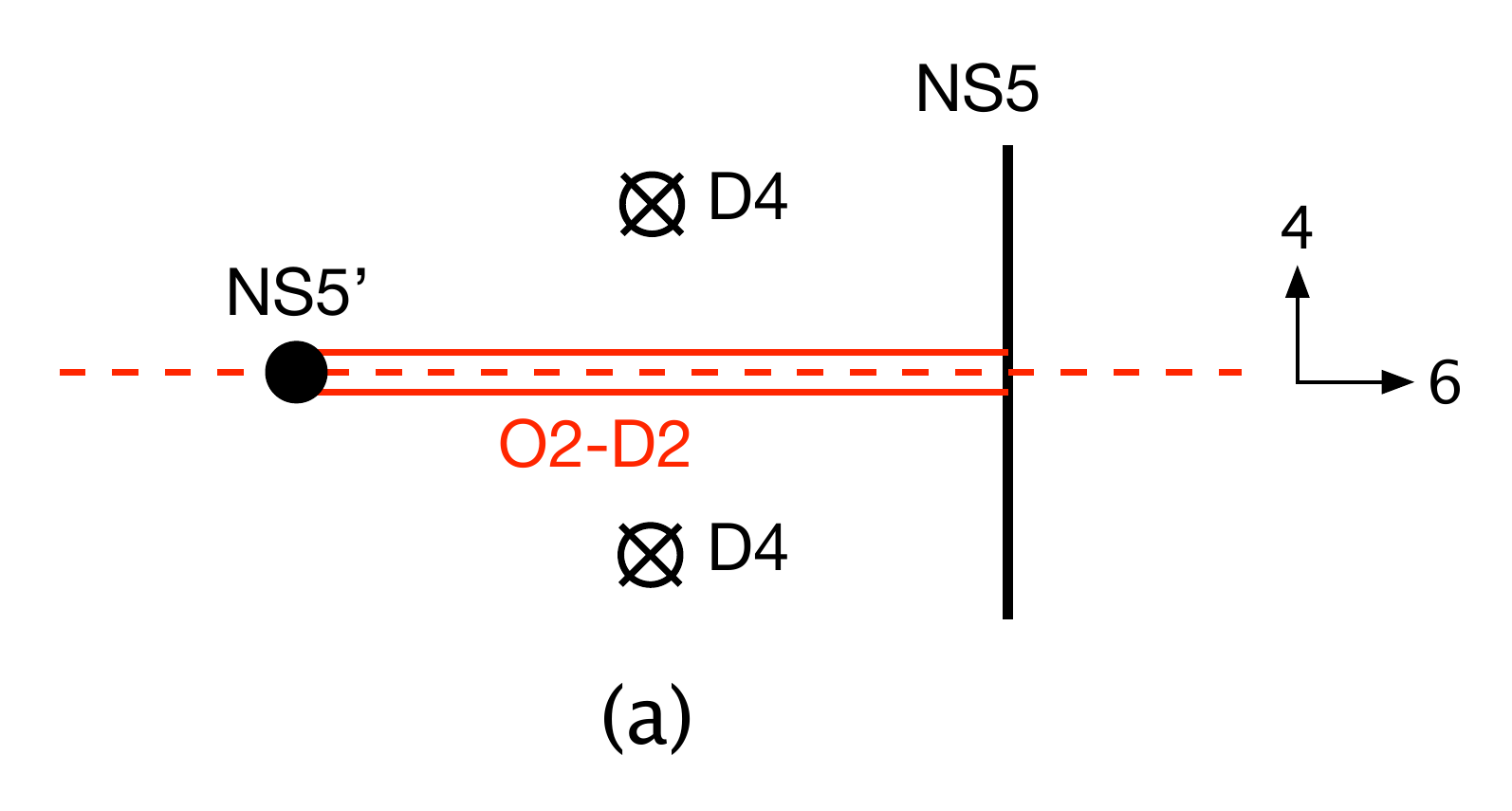} 
\hspace{10pt}
\includegraphics[height=0.2\textwidth]{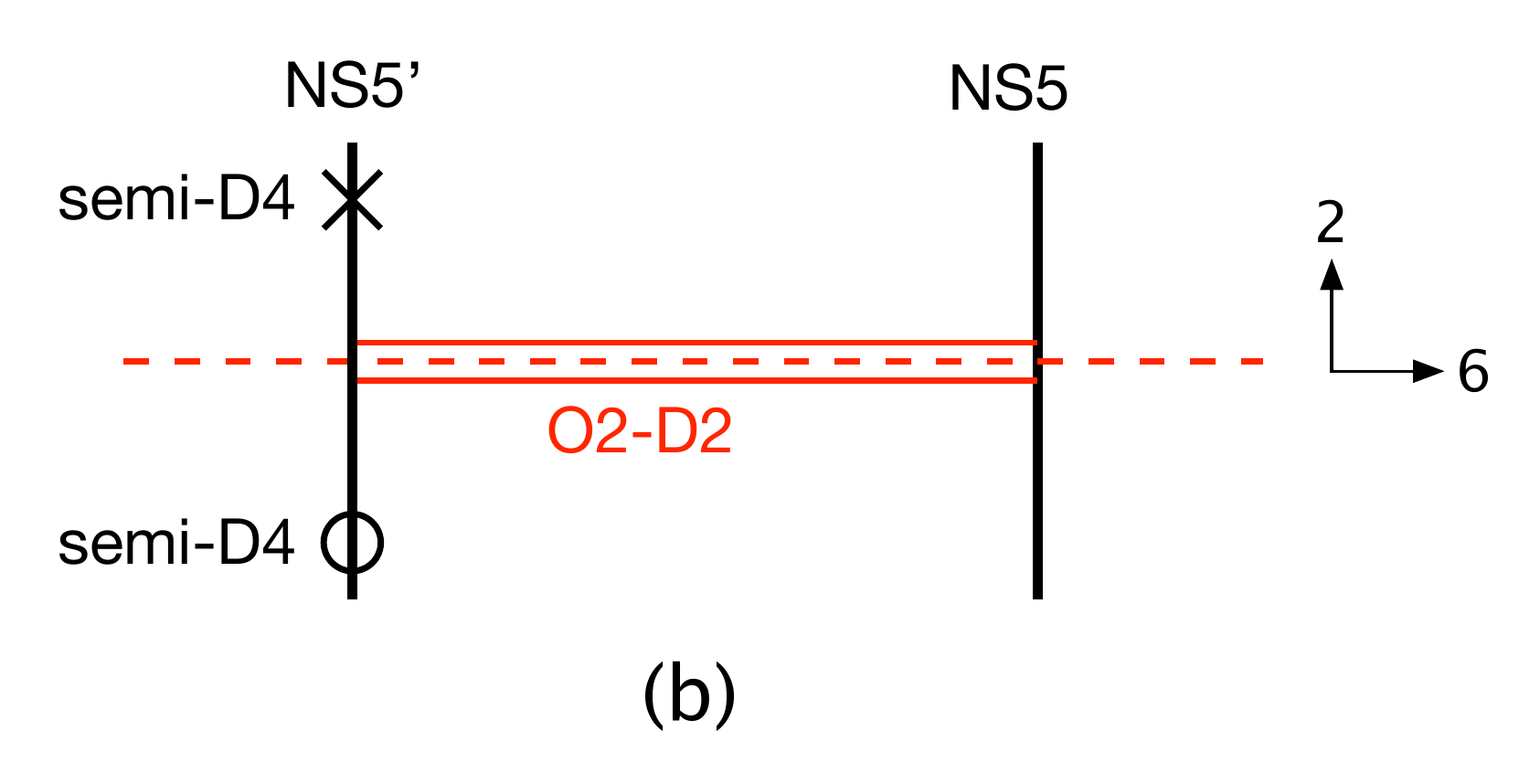}
\caption{(a) Complex mass (b) Twisted mass}
\label{Orientifold2}
\end{figure}

As before, complex masses correspond to the positions of the D4-branes in the $(4,5)$ plane.
However due to the presence of the orientifold plane the D4-branes can only move in pairs in opposite directions, Fig.~\ref{Orientifold2}a.
This is consistent with the $USp(2k)$ theory in which the complex mass matrix is antisymmetric, but it does not
capture the most general symmetric mass matrix of the $O(K)$ theory.
There does not seem to be a brane realization of a more general complex mass deformation in the $O(K)$ theory.
Twisted masses are described by the positions in the (2,3) plane of semi-infinite D4-branes ending on the NS5'-brane, 
Fig.~\ref{Orientifold2}b. There is one parameter per pair of semi-infinite branes which are related by the orientifold projection.
This corresponds to $\tilde{m} = -\hat{m}$ from the point of view of the $U(k)$ theory.

It appears that we have all the ingredients necessary to construct the different theories discussed above,
and to describe their properties.
However this construction leads to three apparent puzzles:
\begin{enumerate}
\item How is the discrete theta parameter $\theta_D$ of the orthogonal theories realized in the brane construction?
\item How are symplectic theories with an odd number of flavors $N=2n+1$ realized?
Inserting an odd number of D4-branes doesn't work, since that changes 
one of the $\mbox{O2}^-$-planes on the outside to an $\widetilde{\mbox{O2}}^-$-plane,
thereby adding another flavor.
So in trying to add one flavor we have really added two.
\item How does the duality move exchanging the two NS5-branes work?
Since the NS5-branes cannot avoid each other in this case (there is no FI parameter),
we must in principle deal with the strong coupling dynamics of their intersection.
The same issue was encountered in the brane construction of ${\cal N}=1$ orthogonal and symplectic theories
in four dimensions \cite{Evans:1997hk}.
In that case it was shown that two additional D4-branes are created when the NS5-branes cross,
explaining the shift by 4 in the duality, for example $SO(k) \leftrightarrow SO(N-k+4)$.
This effect can be accounted for by requiring the conservation of the linking number associated to the NS5-brane \cite{Hanany:1996ie}, 
by including the contribution of the D4-brane charge difference across the NS5-brane.
A similar computation in three dimensions requires the creation of one D3-brane and therefore a shift of 2 in the dualities \cite{Aharony:2008gk}.
However in trying to apply the same logic to the two-dimensional theories, and explain the shift of 1 in the dualities, we encounter a puzzle.
In the absence of D2-branes the linking number of the NS5-brane is $\pm 1/4$, since the charge of the $\mbox{O2}^\pm$-plane is $\pm 1/8$.
The conservation of the linking number therefore requires the creation of a half-D2-brane.
This is impossible. 
In the case of the $\mbox{O2}^+$-plane one cannot have a fractional D2-brane.
It is in principle allowed for an $\mbox{O2}^-$-plane, turning it into an $\widetilde{\mbox{O2}}^-$-plane.
But this is also problematic, since this also seems to require the presence of a D4-brane, which we did not assume.
\end{enumerate}

\noindent These puzzles will be resolved by lifting to M theory.

\subsection{M theory}

The M theory description of the different branes was given in the previous section.
The new ingredient is the O2-plane, which lifts to a pair of OM2-planes located at antipodal points
on the $x^{10}$ circle \cite{Sethi:1998zk,Hanany:2000fq}. 
The OM2-plane is the fixed plane of the orbifold $\mathbb{R}^8/\mathbb{Z}_2$ in M theory.
There are two variants, $\mbox{OM2}^-$ and $\mbox{OM2}^+$, associated to a discrete holonomy of the M theory 3-form.
They carry M2-brane charge $-1/16$ and $3/16$, respectively \cite{Sethi:1998zk}.\footnote{We are using the reduced space normalization of charge.}
The lift of the Type IIA O2-plane is $(\mathbb{R}^7 \times S^1)/\mathbb{Z}_2$, which is two OM2-planes.
The two possibilities for the OM2-plane lead to the four versions of the O2-plane:
\be
\label{O2OM2}
{\mbox O2}^- &=& \mbox{OM}2^- + \mbox{OM}2^- \nonumber \\
\widetilde{\mbox O2}^- &=& \mbox{OM}2^+ + \mbox{OM}2^+ \nonumber \\
{\mbox O2}^+ &=& \mbox{OM}2^- + \mbox{OM}2^+ \nonumber \\
\widetilde{\mbox O2}^+ &=& \mbox{OM}2^+ + \mbox{OM}2^- \nonumber
\ee
In particular this gives the known RR charges of the different O2-planes.
Note that the ${\mbox O2}^+$ and $\widetilde{\mbox O2}^+$ are basically the same object in M theory.
They differ only by a $\pi$ rotation in $x^{10}$.
The transformations of an O2-plane across an NS5-brane or a D4-brane both lift in M theory
to the transformation $\mbox{OM2}^- \rightarrow \mbox{OM2}^+$ across an M5-brane.

In our configuration the positions of the M5-brane and M5'-brane in $x^{10}$ are fixed to the location of an OM2-plane, 
but they can either be on the same OM2-plane or on different OM2-planes.
This will resolve the first two puzzles.
The above choice is a discrete remnant of the continuous degree of freedom that previously described the theta parameter of
the $U(k)$ theory, and will be related to the discrete theta parameter of the $O(K)$ theories, 
and to the choice of an even or odd number of flavors in the $USp(2k)$ theory.

\begin{figure}[h!]
\center
\includegraphics[height=0.25\textwidth]{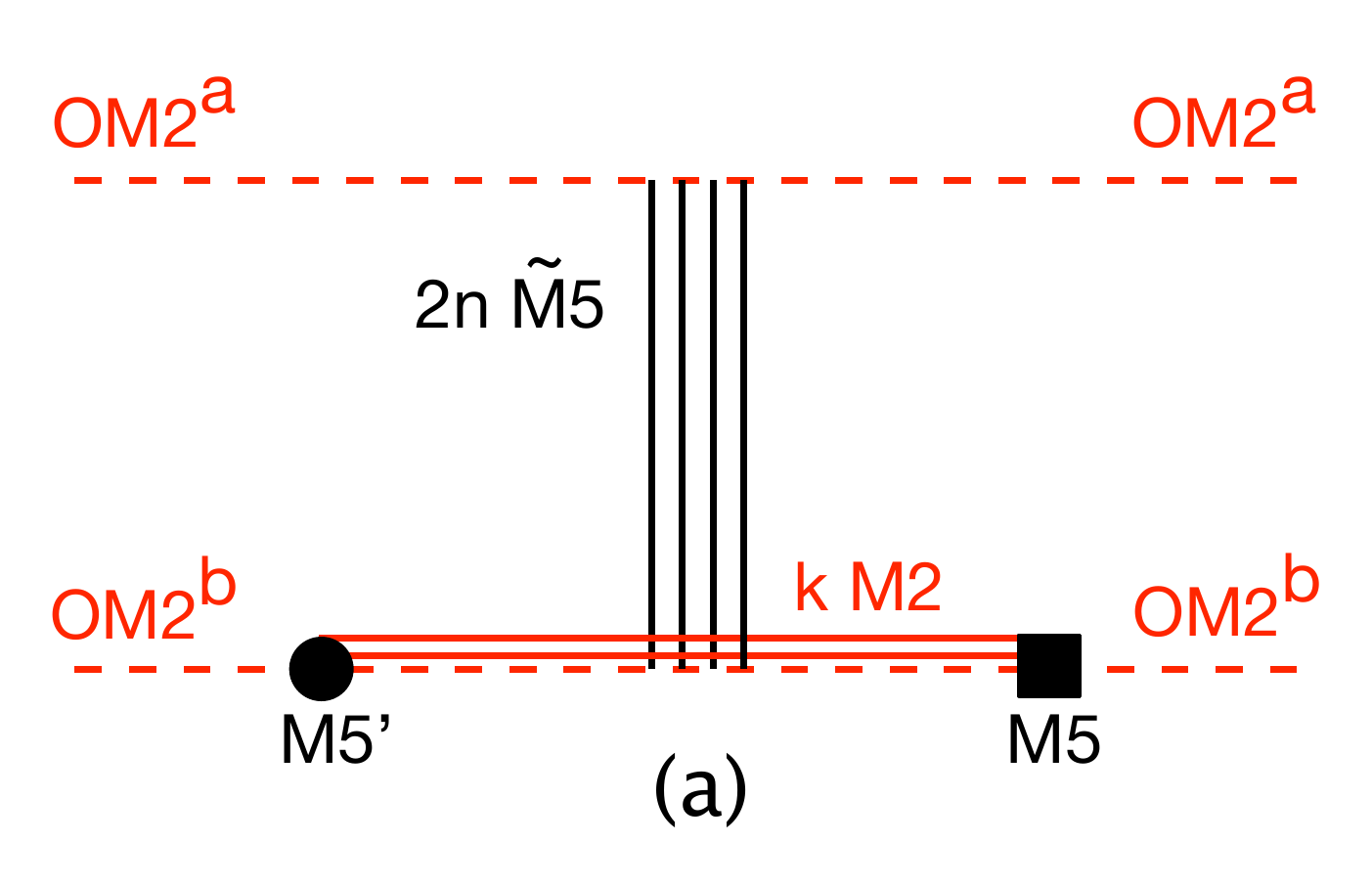} 
\hspace{30pt}
\includegraphics[height=0.25\textwidth]{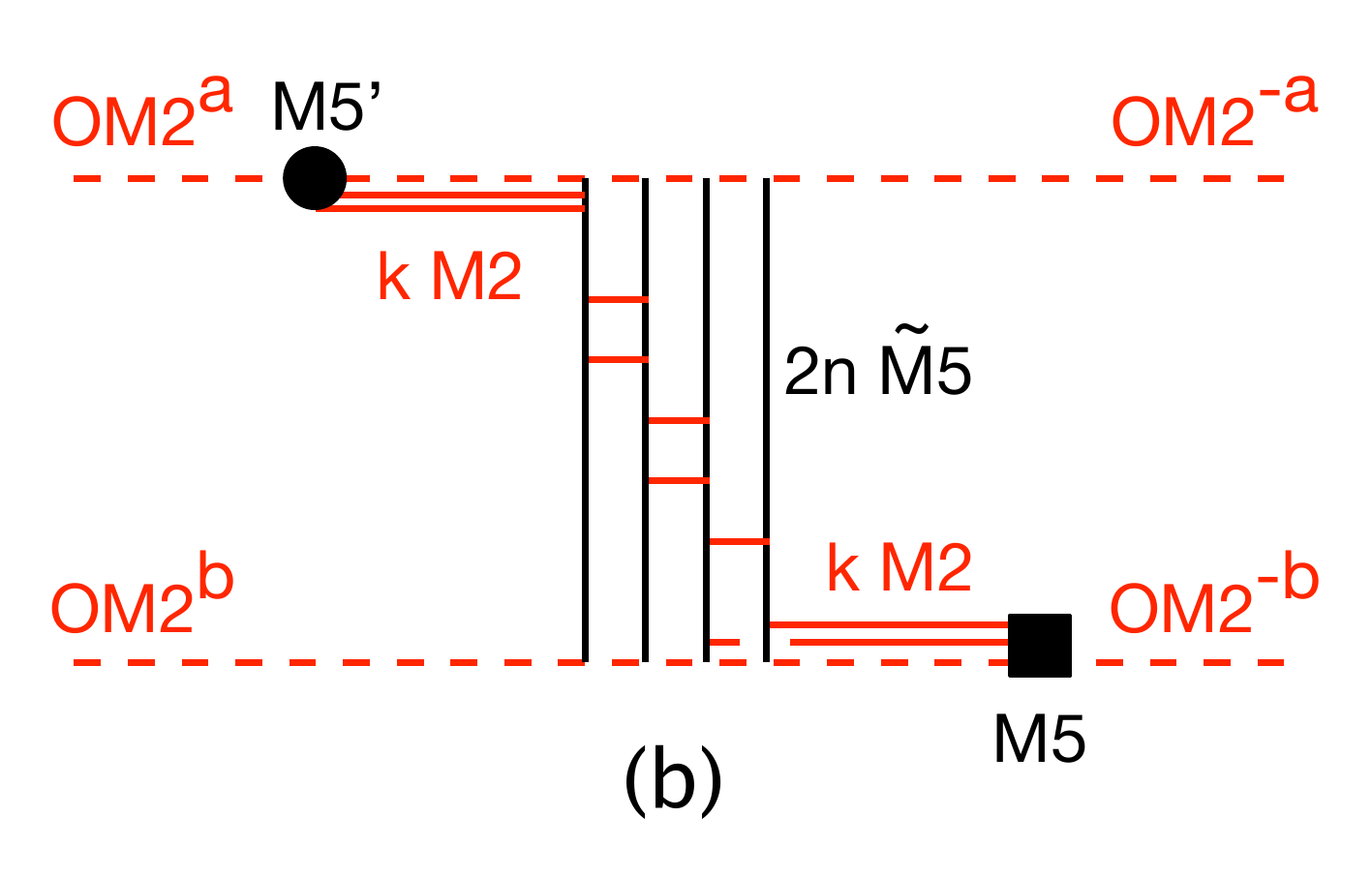}
\caption{M-brane configurations with an even number of flavor branes for $O(K)$ and $USp(2k)$ theories.}
\label{Mbranes1}
\end{figure}

Let us consider first the case of an even number $2n$ of flavor $\widetilde{\mbox{M5}}$-branes.
There are a total of eight inequivalent configurations (see Fig.~\ref{Mbranes1}),
depending on the types of the two OM2-planes on the outside, labeled by $a,b\in \{+,-\}$, and on the relative position of the two M5-branes in $x^{10}$.
The four configurations with the two M5-branes on the same OM2-plane (Fig.~\ref{Mbranes1}a) 
describe 2d gauge theories with a Coulomb branch corresponding to the motion of the M2-branes along the M5-branes
and $\mbox{M5}'$-branes in the (2,3) plane.
The configurations with the M5-branes on different OM2-planes (Fig.~\ref{Mbranes1}b), on the other hand, 
describe theories where the Coulomb branch is lifted, since the M2-branes are forced to break on the 
$\widetilde{\mbox{M5}}$-branes in the vacuum. These describe the {\em regular} theories.

We can now easily identify the gauge theories corresponding to the different choices of $a,b \in \{+,-\}$ using the O2-OM2 relations.
This is summarized in Table~\ref{Theories}.
For example, for $(a,b)=(-,-)$ the configuration in Fig.~\ref{Mbranes1}a describes $USp(2k)$ with $2n$ flavors,
and the configuration in Fig.~\ref{Mbranes1}b describes $USp(2k)$ with $2n+1$ flavors.
The extra flavor in the second case is due to the reduction of the two $\mbox{OM2}^+$-planes on the right to an $\widetilde{\mbox{O2}}^-$-plane.
The other cases include orthogonal theories with an even number of flavors.

\begin{figure}[h!]
\center
\includegraphics[height=0.25\textwidth]{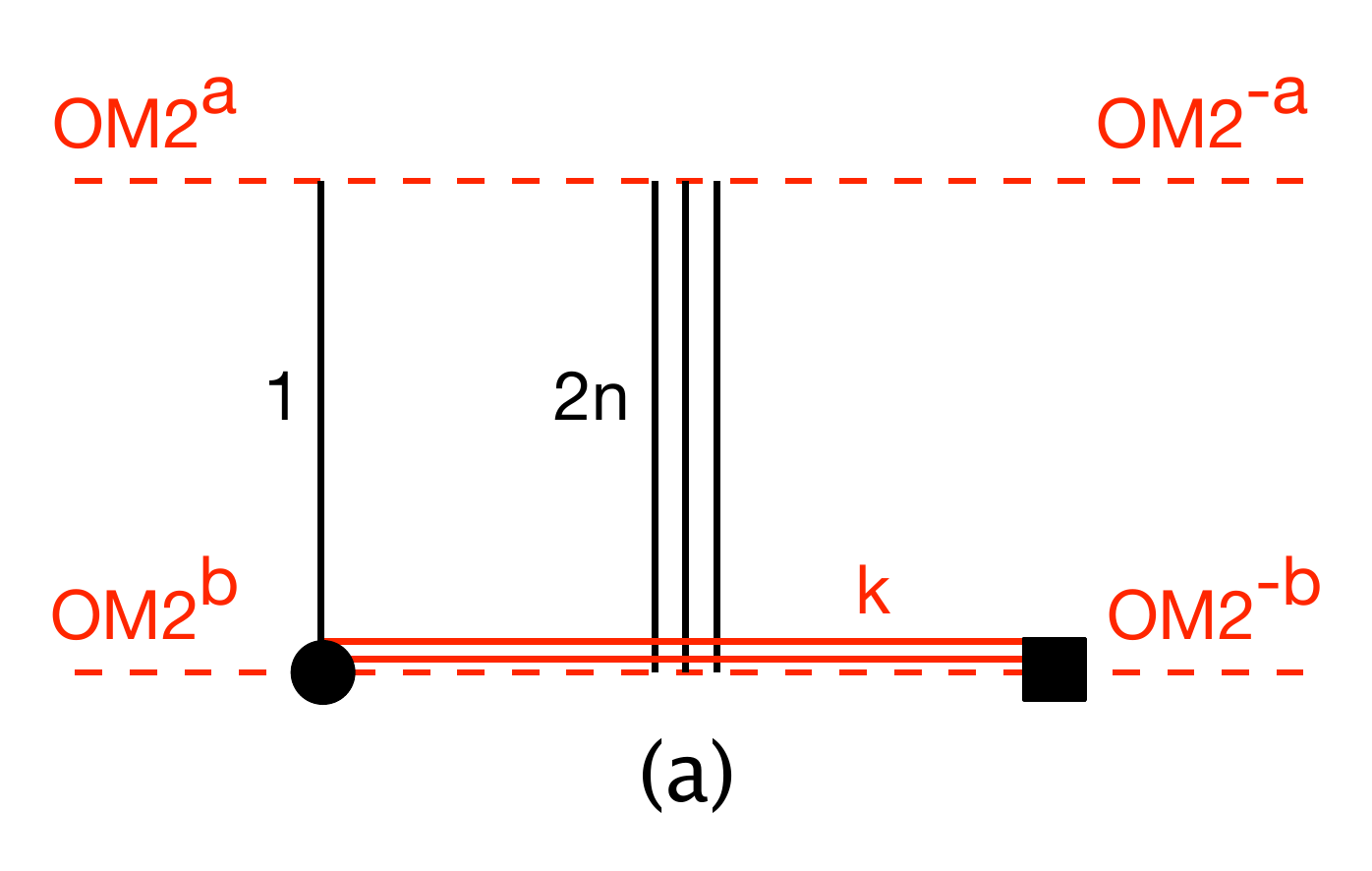} 
\hspace{30pt}
\includegraphics[height=0.25\textwidth]{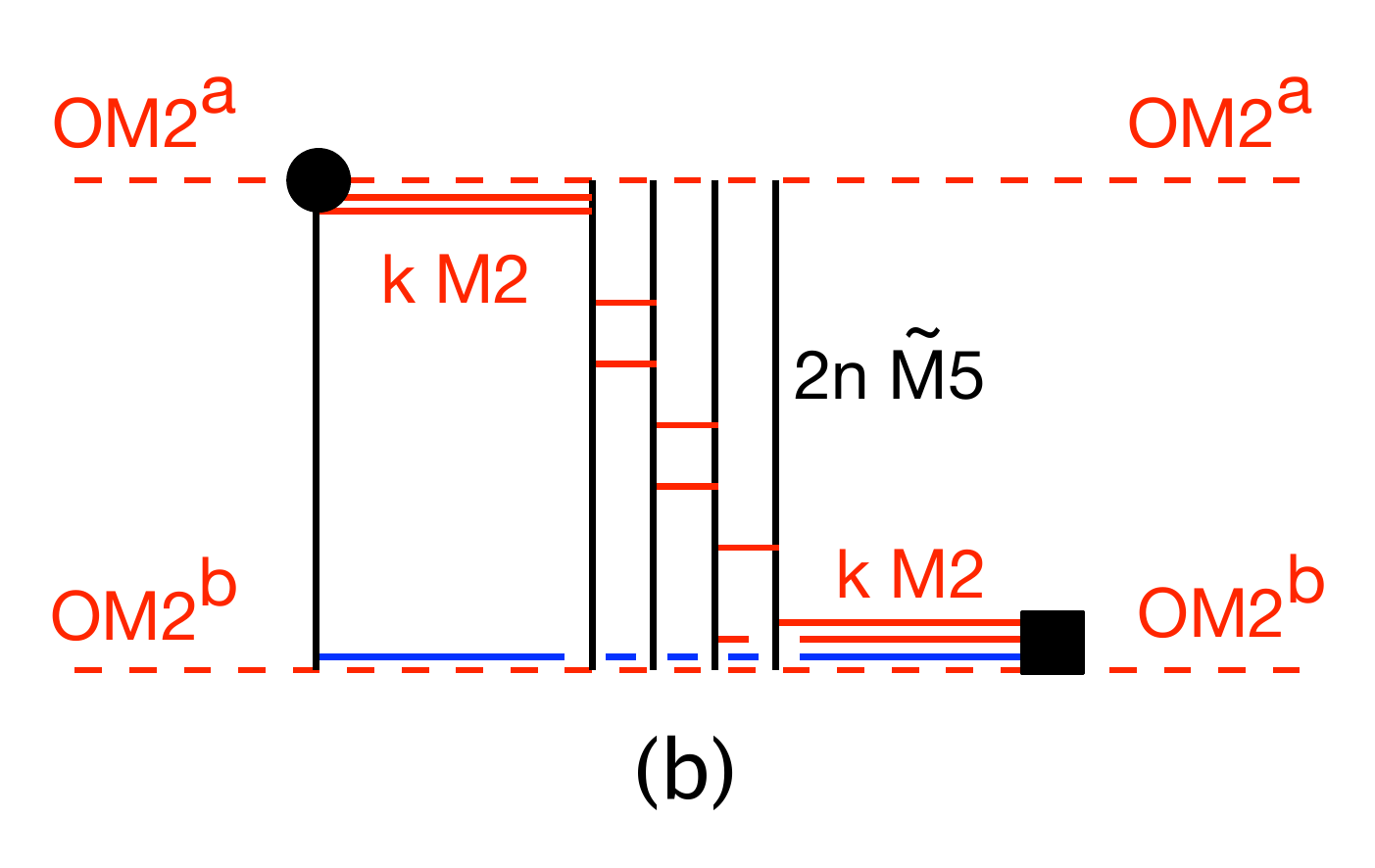}
\caption{M-brane configurations with an odd number of flavor branes for $O(K)$ and $USp(2k)$ theories. 
For $b=+$ a single M2-brane may be suspended between the M5 and the unpaired $\widetilde{\mbox{M5}}$ (blue line).}
\label{Mbranes2}
\end{figure}

To construct orthogonal theories with an odd number of flavors we need to add one more flavor $\widetilde{\mbox{M5}}$-brane, Fig.~\ref{Mbranes2}.
For $(a,b)=(+,-)$ and $(-,+)$ the configurations describe $O(2k)$ and $O(2k+1)$ with an odd number of flavors,
and with $\theta = 0$ or $\pi$.
The unpaired $\widetilde{\mbox{M5}}$-brane is stuck to the $\mbox{M5}'$-brane.
This is the lift of the Type IIA configuration shown in Fig.~\ref{Orientifold1}c.
The other two configurations with $(a,b)=(+,+)$ and $(-,-)$ do not lead to new theories.
These are related to previous configurations by moving the unpaired $\widetilde{\mbox{M5}}$-brane to the left,
giving the $USp(2k)$ theories with an even and odd number of flavors.

\begin{table}[h!]
\begin{center}
\begin{tabular}{|l|l|l|l|}
 \hline 
 $N$ & $(a,b)$ & $\Delta x^{10} = 0$ ({\em irregular}) & $\Delta x^{10}=\pi$ ({\em regular})\\
 \hline 
$2n$ & $(-,-)$ & $USp(2k) + 2n$ & $USp(2k) + 2n + 1$ \\
& $(+,+)$ & $USp(2k) + 2n + 2$ & $USp(2k) + 2n + 1$ \\
& $(+,-)$ & $O(2k+1)_{\pi} + 2n$ & $O(2k)_{\pi} + 2n $\\
& $(-,+)$ & $O(2k)_{0} + 2n$ & $O(2k+1)_{0} + 2n$\\
 \hline
 $2n+1$ & $(-,-)$ & $USp(2k) + 2n+2$ & $USp(2k) + 2n + 1$ \\
& $(+,+)$ & $USp(2k) + 2n + 2$ & $USp(2k) + 2n + 3$ \\
& $(+,-)$ & $O(2k)_{\pi} + 2n + 1$ & $O(2k+1)_{\pi} + 2n + 1$\\
& $(-,+)$ & $O(2k+1)_{0} + 2n + 1$ & $O(2k)_{0} + 2n + 1$\\
\hline
 \end{tabular}
 \end{center}
\caption{Theories described by the OM2 M-brane configurations.}
\label{Theories}
\end{table}

\begin{figure}[h!]
\center
\includegraphics[height=0.2\textwidth]{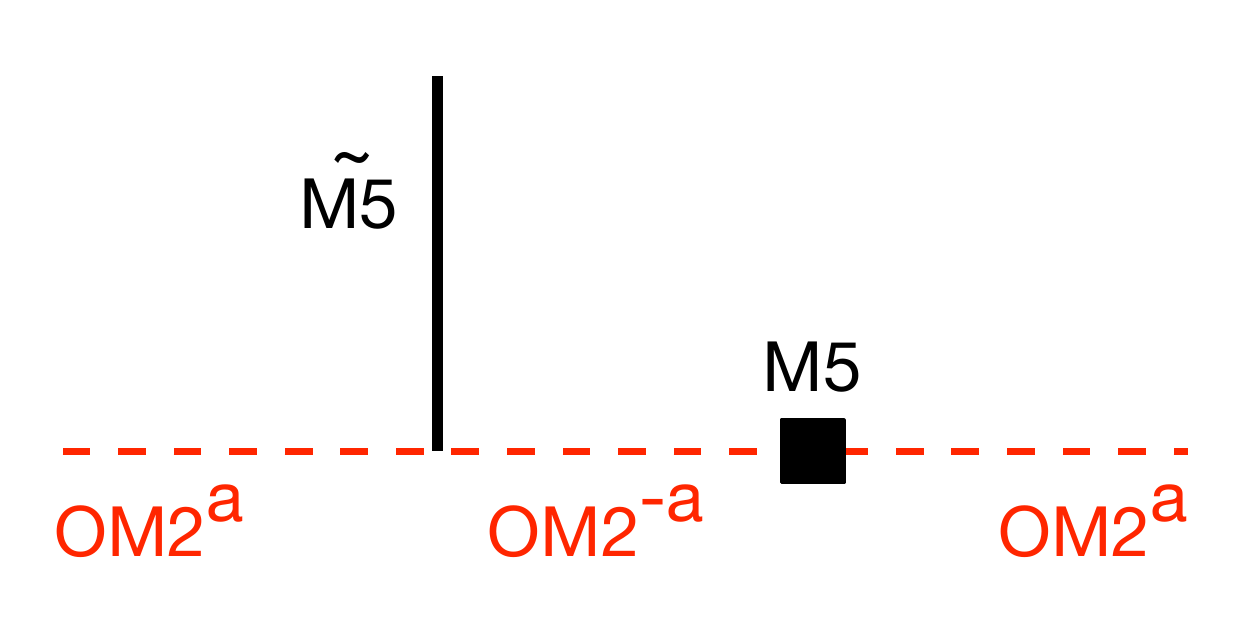} 
\hspace{30pt}
\includegraphics[height=0.2\textwidth]{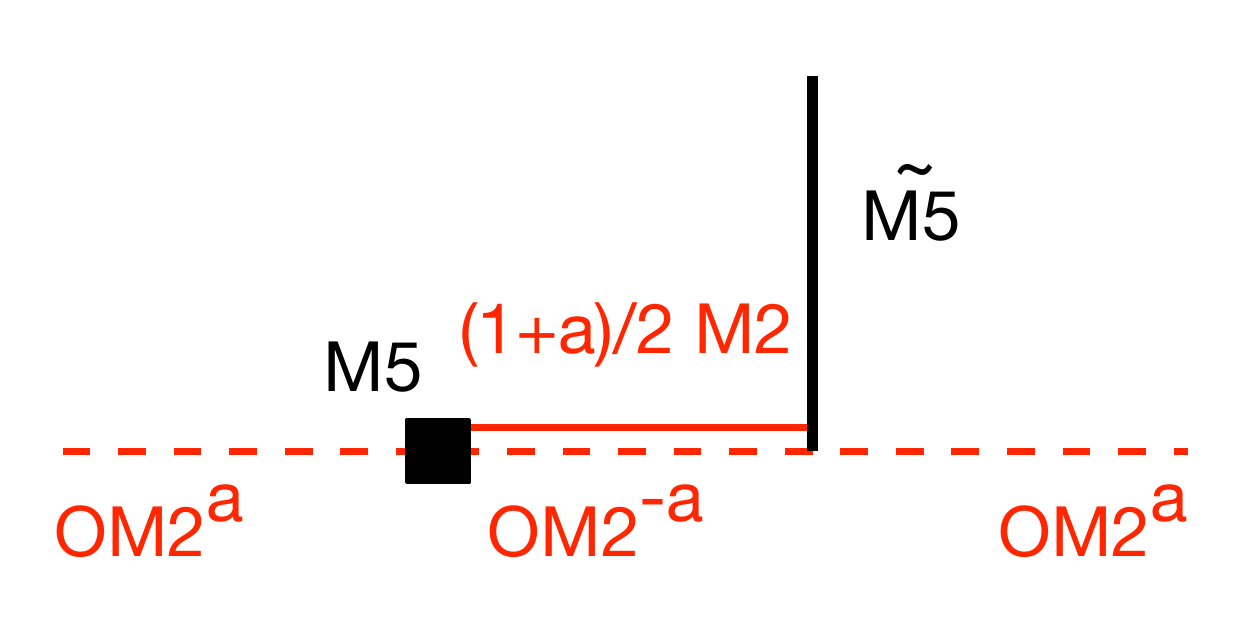}
\caption{Creation or not of an M2-brane when M5-branes cross on an OM2-plane.}
\label{M5exchange}
\end{figure}

Counting the number of supersymmetric vacua requires a generalization of the s-rule to the M theory configurations.
Since this is intimately related to the brane creation phenomenon, which will be important also in the description of the
dualities, we pause momentarily to discuss this.
The basic situation we want to consider is an M5-brane and an $\widetilde{\mbox{M5}}$-brane on an OM2-plane, Fig.~\ref{M5exchange}.
We know that in flat space, exchanging these two M5-branes leads to the creation of a single M2-brane.
Naively one might conclude that a half-M2-brane should be created when the two ``fractional" M5-branes are exchanged on the OM2-plane.
But this is not really possible due to the charge quantization condition in M theory \cite{Witten:1996md}.
To see what actually happens we will use the linking number argument.
The linking number associated to the M5-brane is given by
\be
L_{M5} = \frac{1}{2}(N^R_{\widetilde{M5}} - N^L_{\widetilde{M5}}) + N^L_{M2} - N^R_{M2} \,,
\ee
where the $\mbox{OM2}$-plane contributes its M2-brane charge as well, which for the $\mbox{OM2}^a$-plane is $(1+2a)/16$.
In exchanging the two M5-branes the linking number changes by
\be
\Delta L_{M5}  = \mbox{} -\frac{1}{4} - \frac{a}{4} - \left(\frac{1}{4} + \frac{a}{4}\right) = \mbox{} - \frac{1+a}{2} \,.
\ee
Therefore a single M2-brane must be created for $a=+$, and none for $a=-$, Fig.~\ref{M5exchange}b.
Iterating the process for $N$ $\widetilde{\mbox{M5}}$-branes we find that $n$ M2-branes are created if $N=2n$,
and $n + \frac{1+a}{2}$ M2-branes are created if $N=2n+1$.
We can also conclude from this the form of s-rule in this situation.
For an even number of $\widetilde{\mbox{M5}}$-branes, there can be at most one full M2-brane per pair.
In the covering space there are two M2-branes, one ending on one $\widetilde{\mbox{M5}}$-brane and the other on its partner.
For an odd number $N=2n+1$ of $\widetilde{\mbox{M5}}$-branes, the same rule applies to $2n$ of them.
For the remaining unpaired $\widetilde{\mbox{M5}}$-brane, $N_{M2}\leq 1$ if $a=+$ and $N_{M2}=0$ if $a=-$.

Coming back to the counting problem, we find that for the configurations with $N=2n$ supersymmetry is unbroken if 
$n\geq k$, and there are $n\choose k$ supersymmetric vacua.
For the configurations with $N=2n+1$ the condition for unbroken supersymmetry is $n\geq k-1$ for $b=+$,
and $n\geq k$ for $b=-$.
For $b=-$ the counting is the same as before since we cannot suspend an M2-brane between the M5-brane and the unpaired
$\widetilde{\mbox{M5}}$-brane, and we again get $n\choose k$ vacua.
For $b=+$ we also have to add the configurations where one M2-brane connects the M5-brane to the 
unpaired $\widetilde{\mbox{M5}}$-brane (see Fig.~\ref{Mbranes2}b).
This gives ${n\choose k} + {n\choose k-1} = {n+1\choose k}$ supersymmetric vacua. 
Modulo the precise identification of the orthogonal theories as $O(K)_+$, $O(K)_-$, or $SO(K)$, which we are not able to make,
the counting of supersymmetric ground states agrees with the field theory results in \cite{Hori:2011pd} (see Table (4.20) there).

\subsection{Duality moves}

\begin{figure}[h!]
\center
\includegraphics[height=0.16\textwidth]{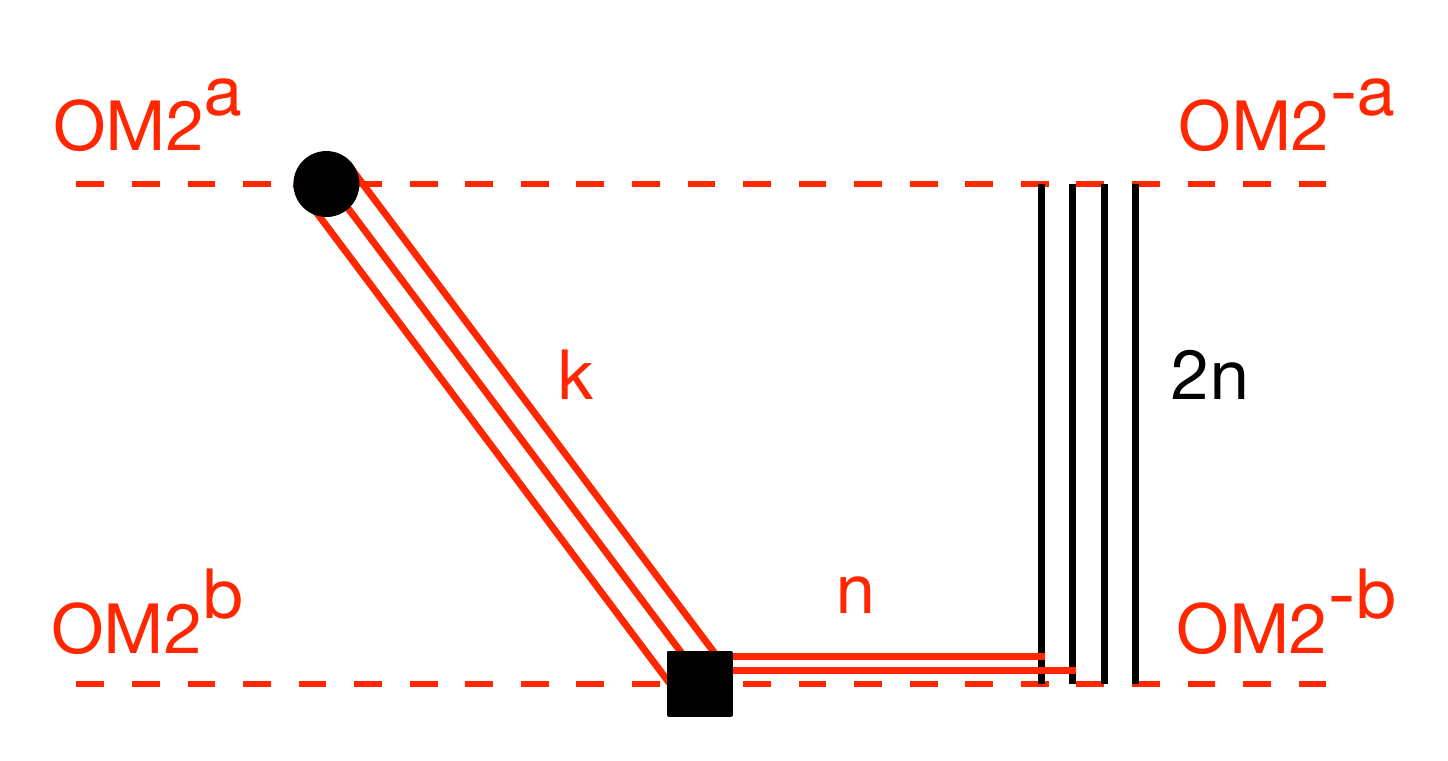} 
\hspace{5pt}
\includegraphics[height=0.16\textwidth]{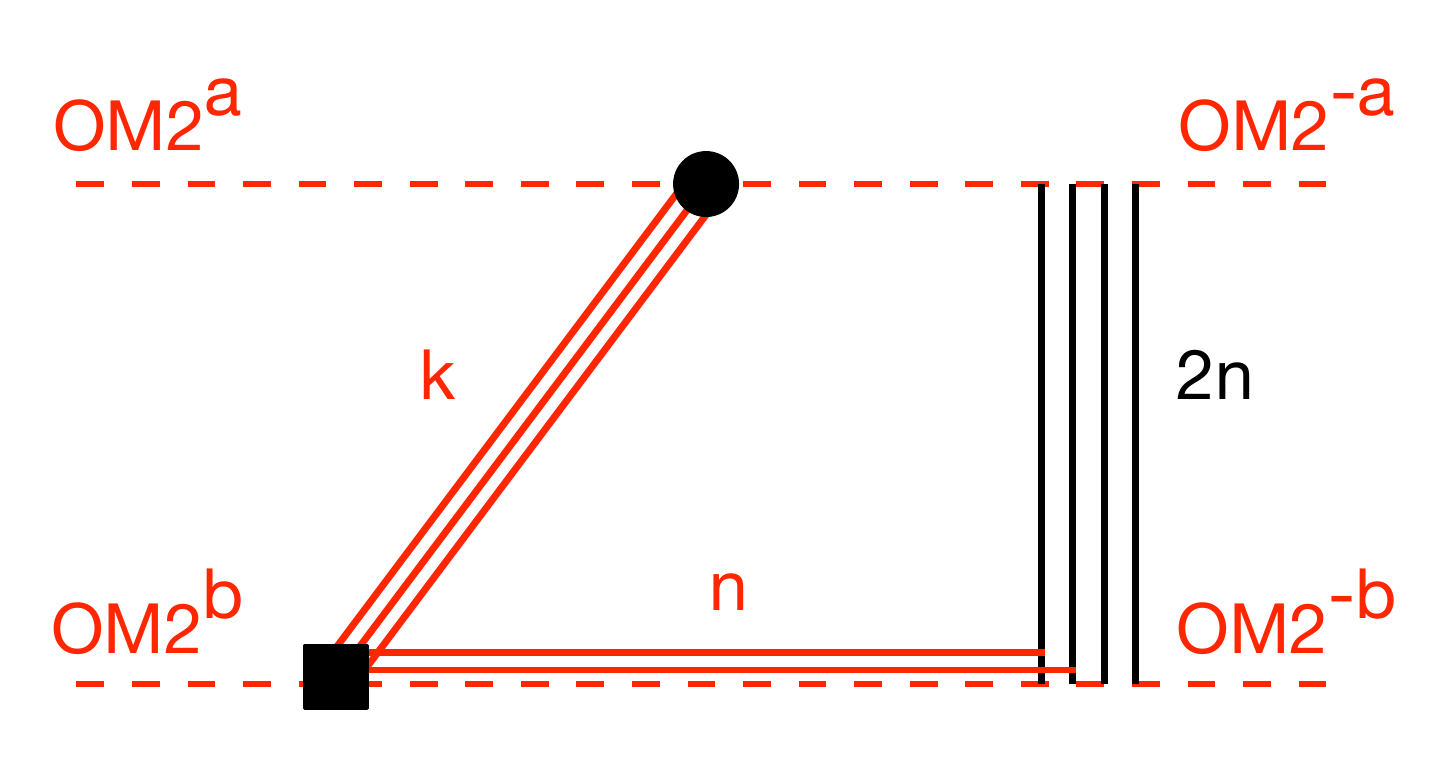}
\hspace{5pt}
\includegraphics[height=0.16\textwidth]{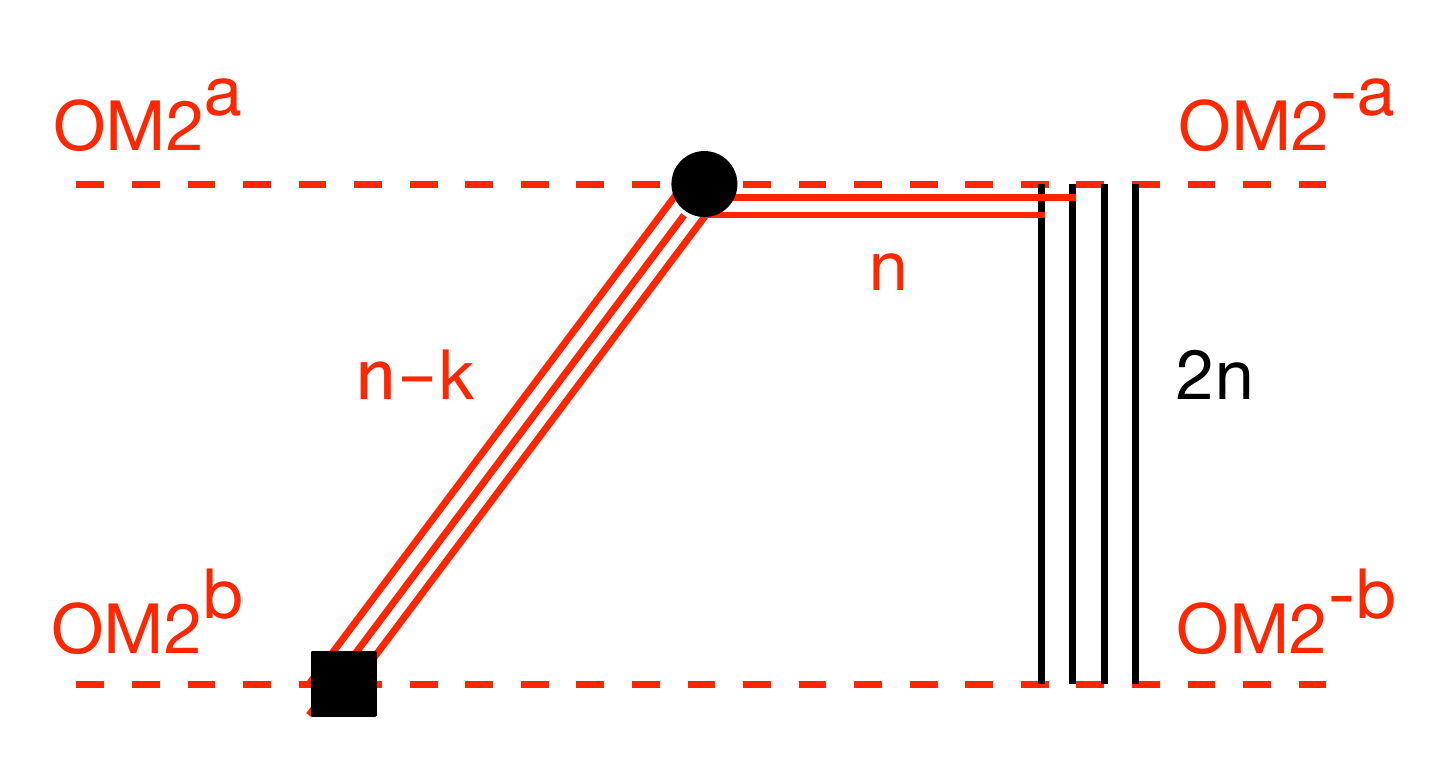} 
\caption{The duality move in M theory for $N=2n$.}
\label{Mbranes3}
\end{figure}

\begin{figure}[h!]
\center
\includegraphics[height=0.16\textwidth]{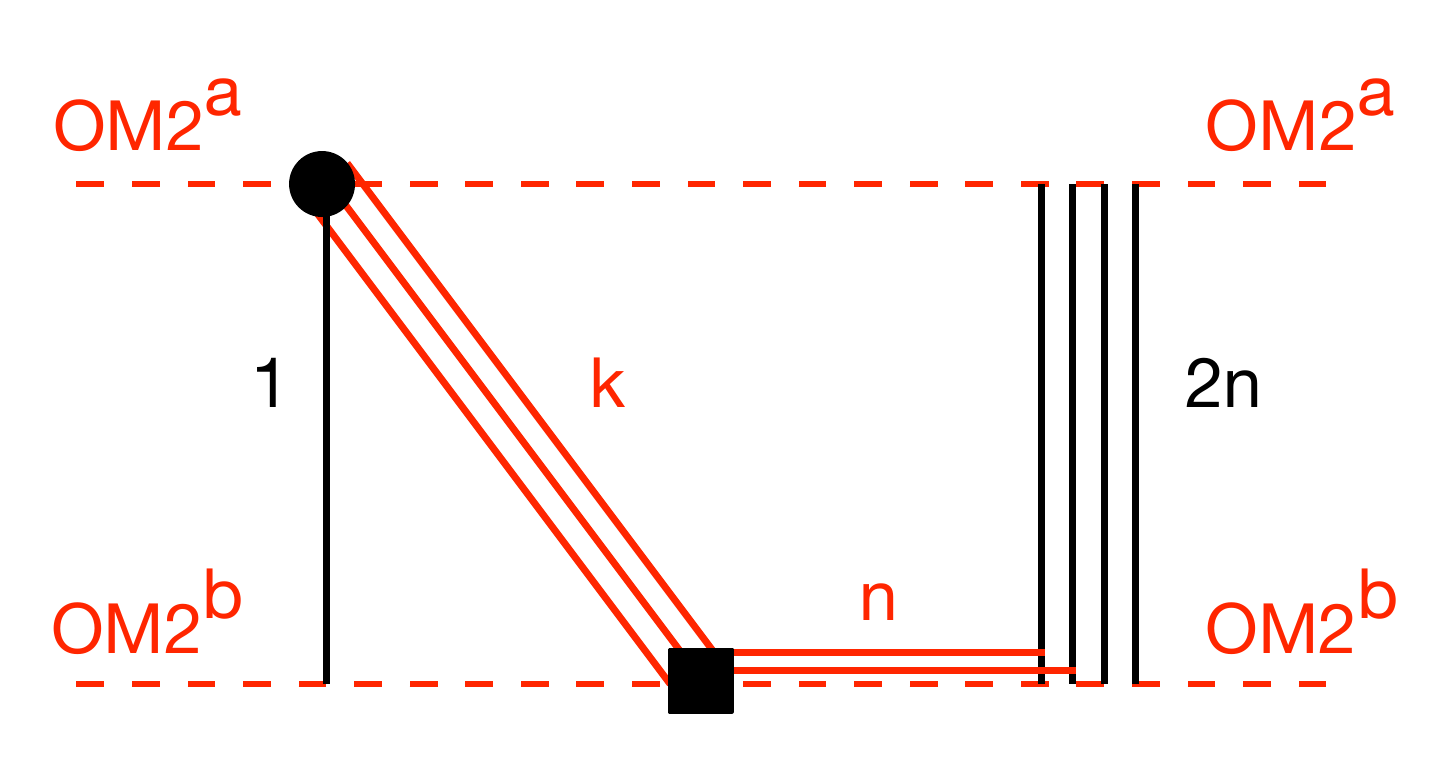} 
\hspace{5pt}
\includegraphics[height=0.16\textwidth]{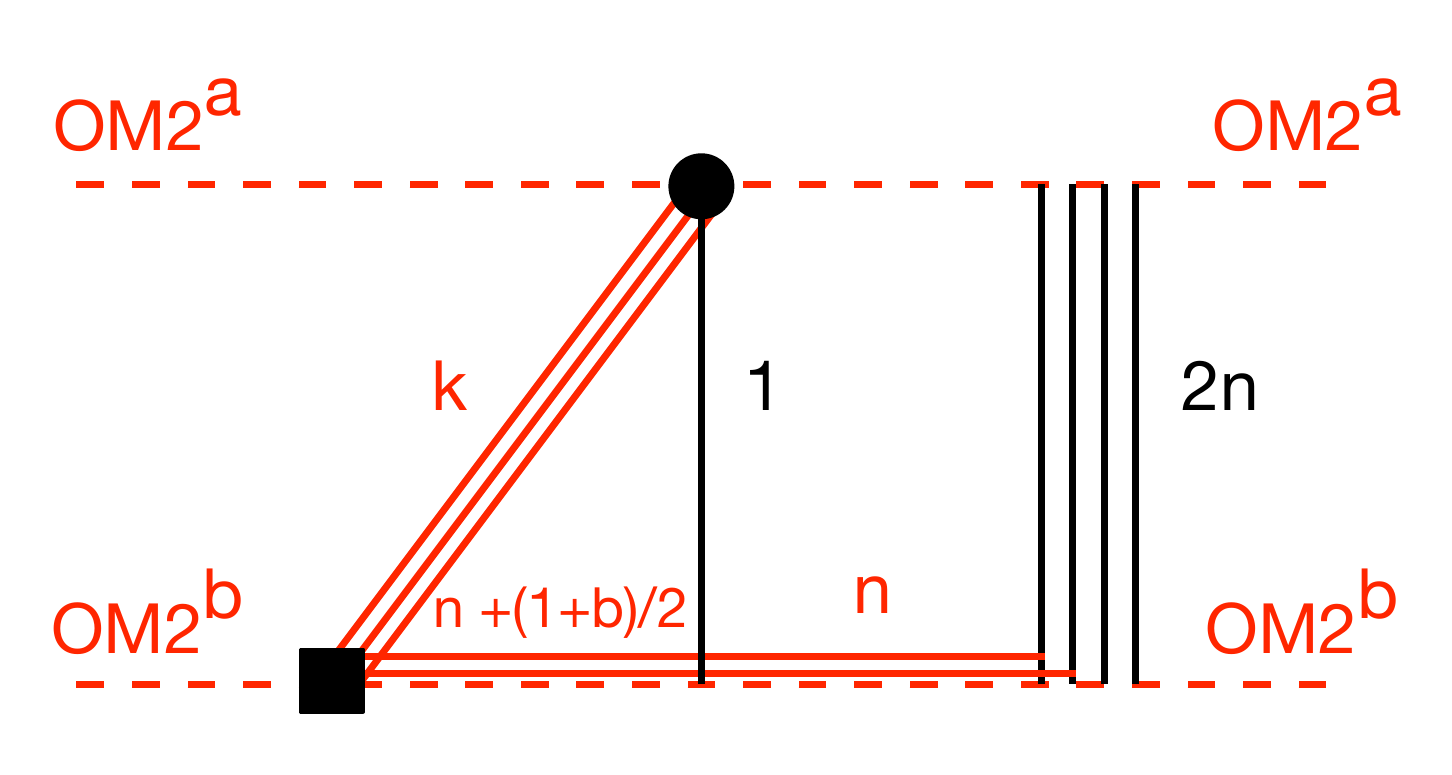}
\hspace{5pt}
\includegraphics[height=0.16\textwidth]{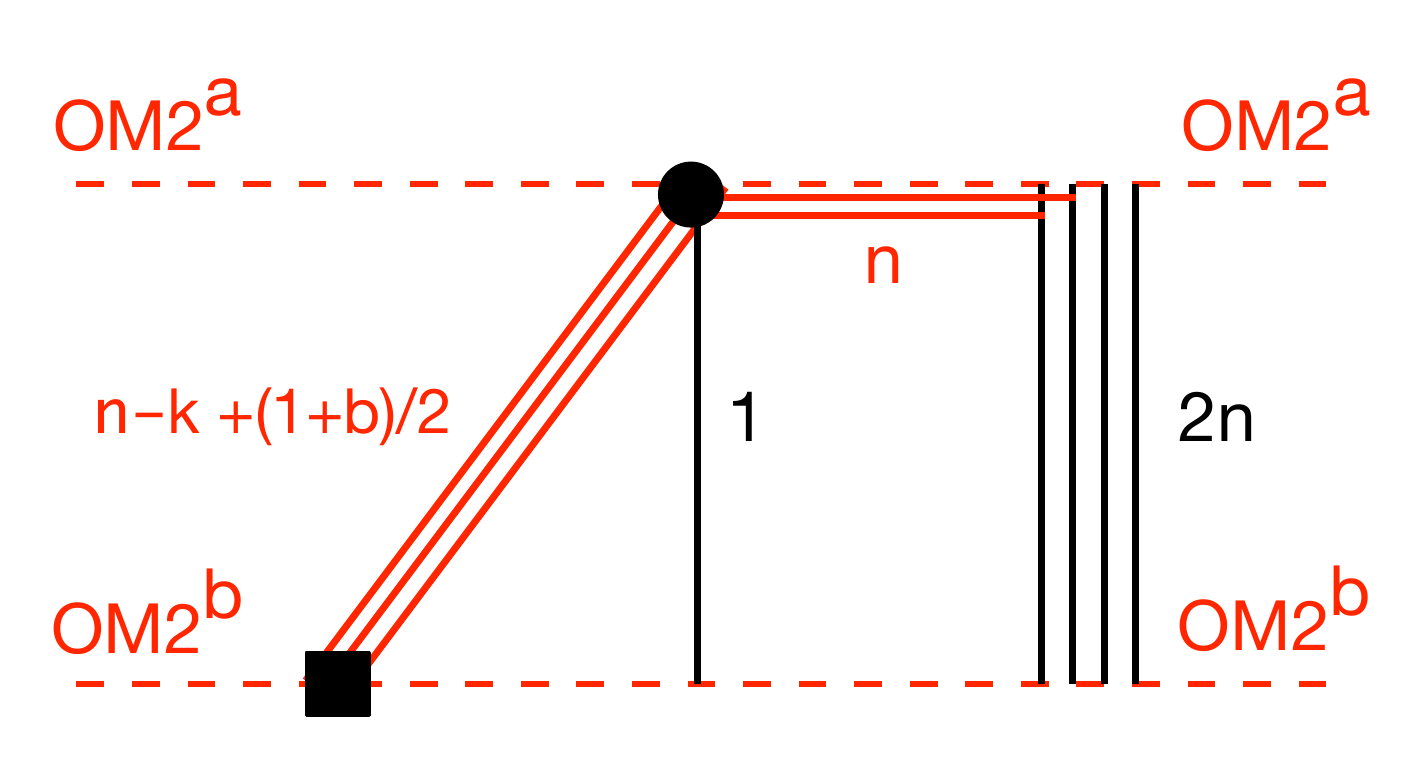} 
\caption{The duality move in M theory for $N=2n+1$.}
\label{Mbranes4}
\end{figure}

The final puzzle in the Type IIA construction is related to the duality move exchanging the NS5-brane and NS5'-brane.
This issue is resolved in the M theory configuration of the {\em regular} theories, since the M5-brane and M5'-brane never intersect.
The duality moves are shown for $N=2n$ in Fig.~\ref{Mbranes3}, and for $N=2n+1$ in Fig.~\ref{Mbranes4}.
We first move $2n$ of the $\widetilde{\mbox{M5}}$-branes to the right and across the M5-brane,
which leads to the creation of $n$ M2-branes. Then we exchange the M5-brane and M5'-brane in the $x^6$ direction,
which leads to the creation of an additional M2-brane in the odd $N$ case if $b=+$.
Finally we recombine M2-branes to minimize the energy.

Using the O2-OM2 relations we can then identify the dual theories.
From Fig.~\ref{Mbranes3} with $(a,b) = (+,+)$ or $(-,-)$ we get the duality for the symplectic theory with an odd number of flavors,
\be
\label{USpDuality}
USp(2k) + 2n +1 \longleftrightarrow USp(2n-2k) +2n +1 \,,
\ee
and with $(a,b)=(+,-),(-,+)$ we get the dualities for the orthogonal theories with an even number of flavors,
\be
O(2k) + 2n  & \longleftrightarrow & O(2n-2k+1) + 2n \\
O(2k+1) + 2n  & \longleftrightarrow & O(2n-2k) + 2n \,.
\ee
Fig.~\ref{Mbranes4} with $(a,b)=(-,+),(+,-)$ shows the duality for the orthogonal theories with an odd number of flavors,
\be
O(2k) + 2n+1 & \longleftrightarrow & O(2n-2k+2) + 2n+1 \\
O(2k+1) + 2n+1 & \longleftrightarrow & O(2n-2k+1) + 2n+1 \,.
\label{ODualityOddOdd}
\ee
These are all in agreement with the proposed dualities, again
modulo the identification of the orthogonal theories as $O_+$, $O_-$, or $SO$.

\subsection{Back to Type IIA}

The crucial observation that led to the resolution of the puzzles we encountered in the Type IIA brane construction
was that in the lift to M theory we had a choice of putting the M5-brane and M5'-brane on the same OM2-plane or on different OM2-planes.
It is actually instructive to reduce back to Type IIA  string theory and interpret this observation in the Type IIA brane construction.

In flat space, the position of the M5-brane in $x^{10}$ corresponds to a uniform VEV of the compact scalar in the
NS5-brane tensor multiplet. 
On the other hand an M5-brane wrapping $x^{10}$ corresponds to a D4-brane.
More generally a non-uniform VEV describes an NS5-D4 bound state.
For example, assume that $x^{10}  = \ell x^1$ (and that $x^1$ is compact).
This describes an M5-brane wrapping $x^{10}$ $\ell$ times as it wraps $x^1$ once,
which corresponds to a bound state of one NS5-brane and $\ell$ D4-branes.
From the point of view of the NS5-brane this is seen in the worldvolume coupling
\be
\int_{\mathbb{R}^6} C_5 \wedge dx^{10} \,.
\ee

The compact scalar field $x^{10}$ is odd under the orientifold projection of the O2-plane,
but a discrete remnant taking values in $\{0,\pi\}$ remains.
The non-trivial class corresponds to an NS5-D4 bound state.
In other words the reduction of a single M5-brane to Type IIA string theory depends on its
discrete $x^{10}$ position. For $x^{10}=0$ it reduces to an NS5-brane, and for $x^{10}=\pi$ it reduces to 
an NS5-D4 bound state.

There is another way to interpret this bound state.
Consider the orthogonal theory with a single flavor, and turn on a twisted mass $\tilde{m}$.
In the brane construction this is described by breaking the flavor D4-brane on the NS5'-brane
and separating the two halves symmetrically to $\sigma = \pm \tilde{m}$, as in Fig.~\ref{Orientifold2}b.
As $\tilde{m}\rightarrow\infty$ this gives an NS5'-D4 bound state.
On the other hand from the point of view of the 2d orthogonal gauge theory this shifts $\theta_D \rightarrow \theta_D + \pi$.
This is consistent with the identification of $\theta_D$ with the discrete relative $x^{10}$ position of the M5-branes.

\begin{figure}[h!]
\center
\includegraphics[height=0.18\textwidth]{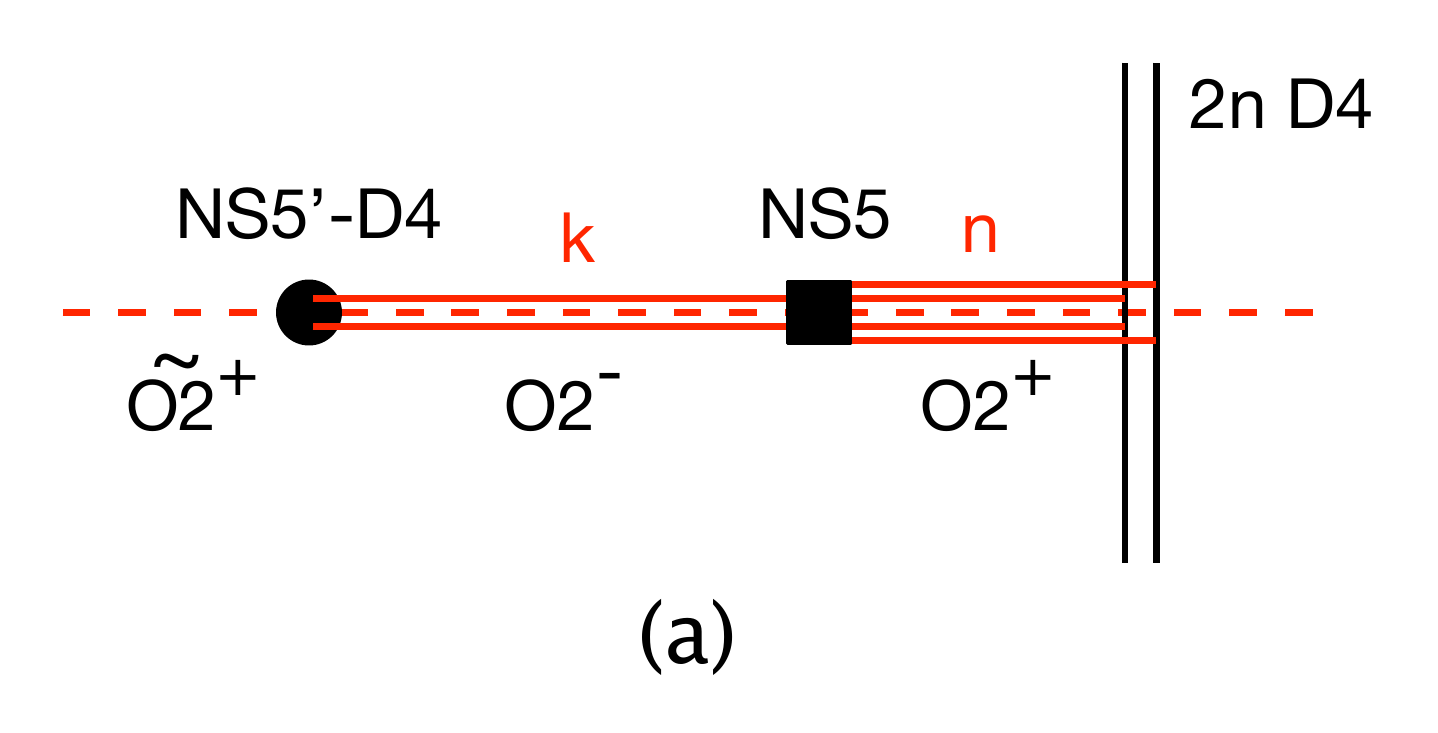} 
\hspace{30pt}
\includegraphics[height=0.18\textwidth]{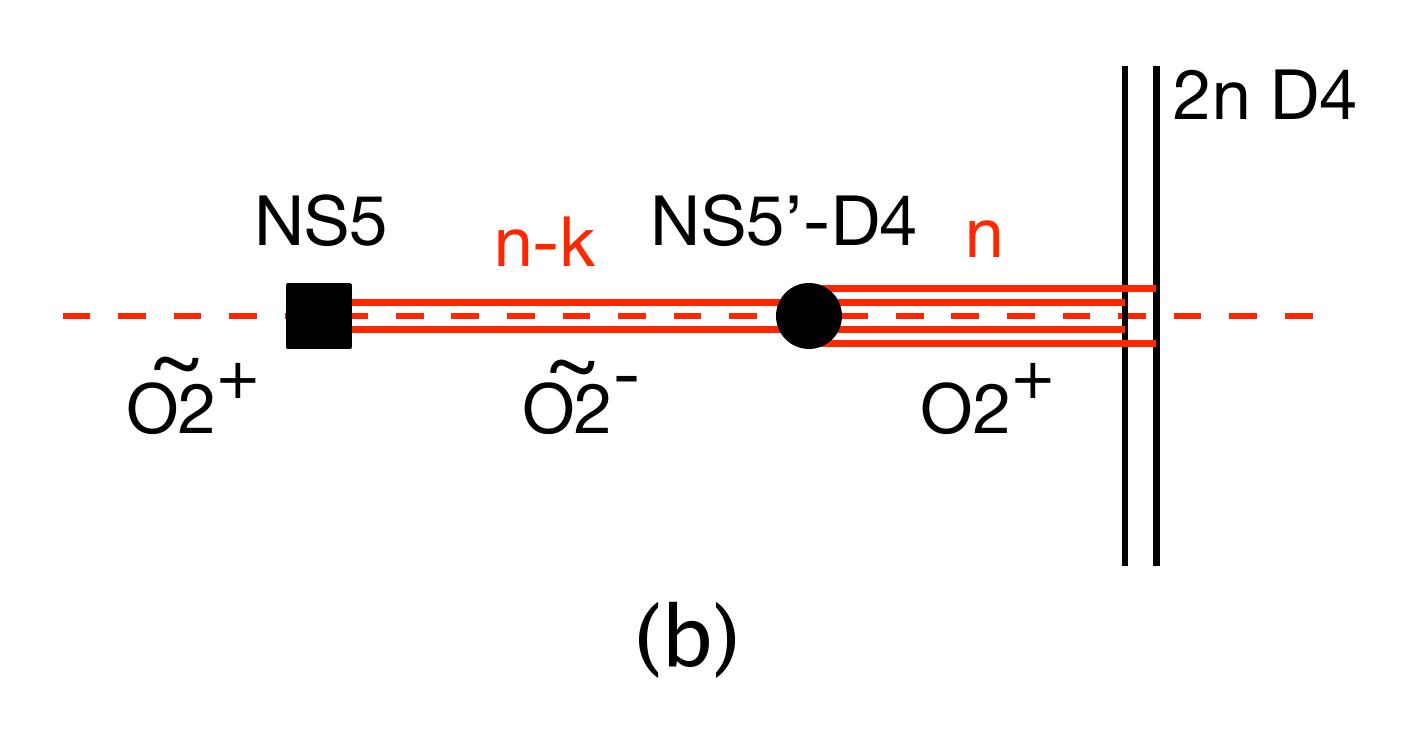}
\caption{Duality move in the Type IIA brane configuration for $O(2k)_\pi + 2n$.}
\label{Orientifold3}
\end{figure}

\begin{figure}[h!]
\center
\includegraphics[height=0.18\textwidth]{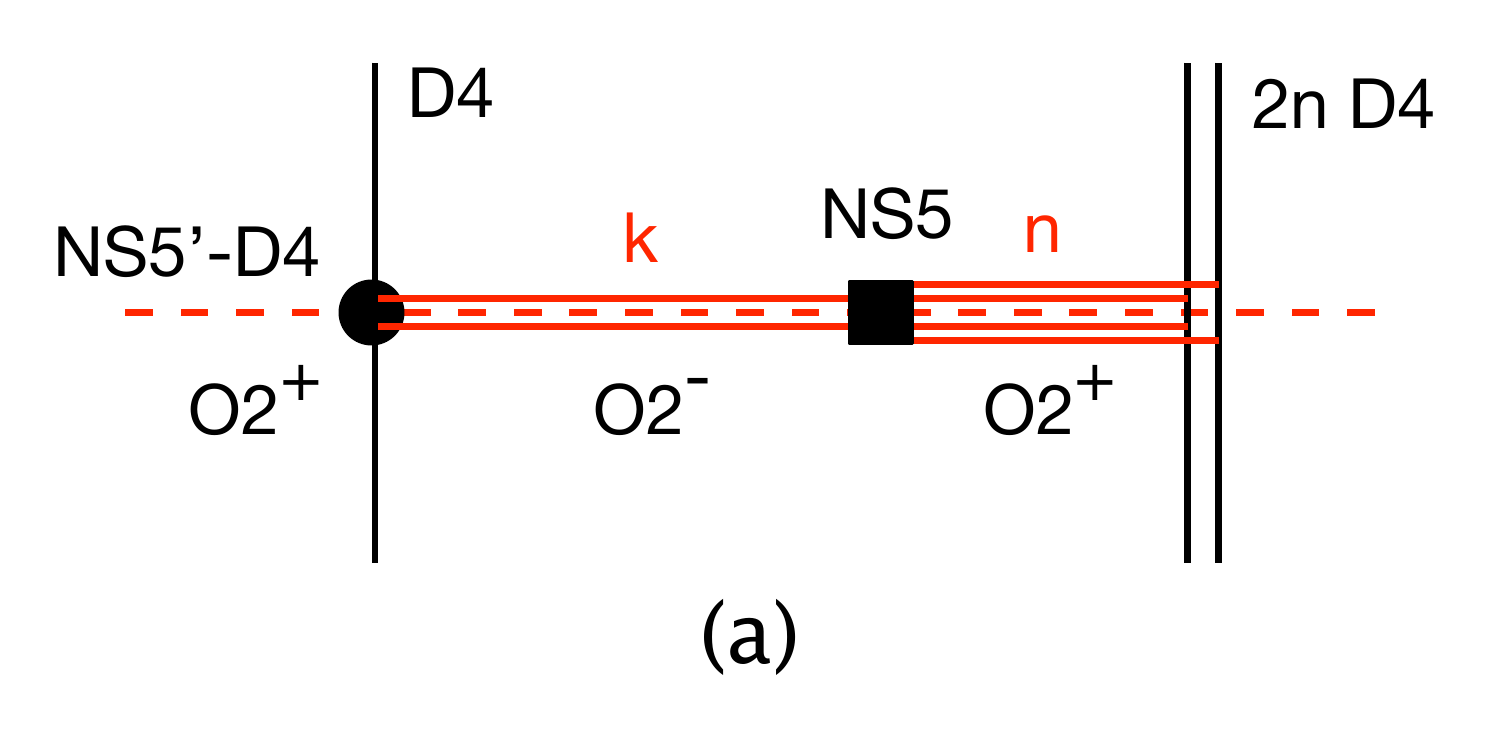} 
\hspace{30pt}
\includegraphics[height=0.18\textwidth]{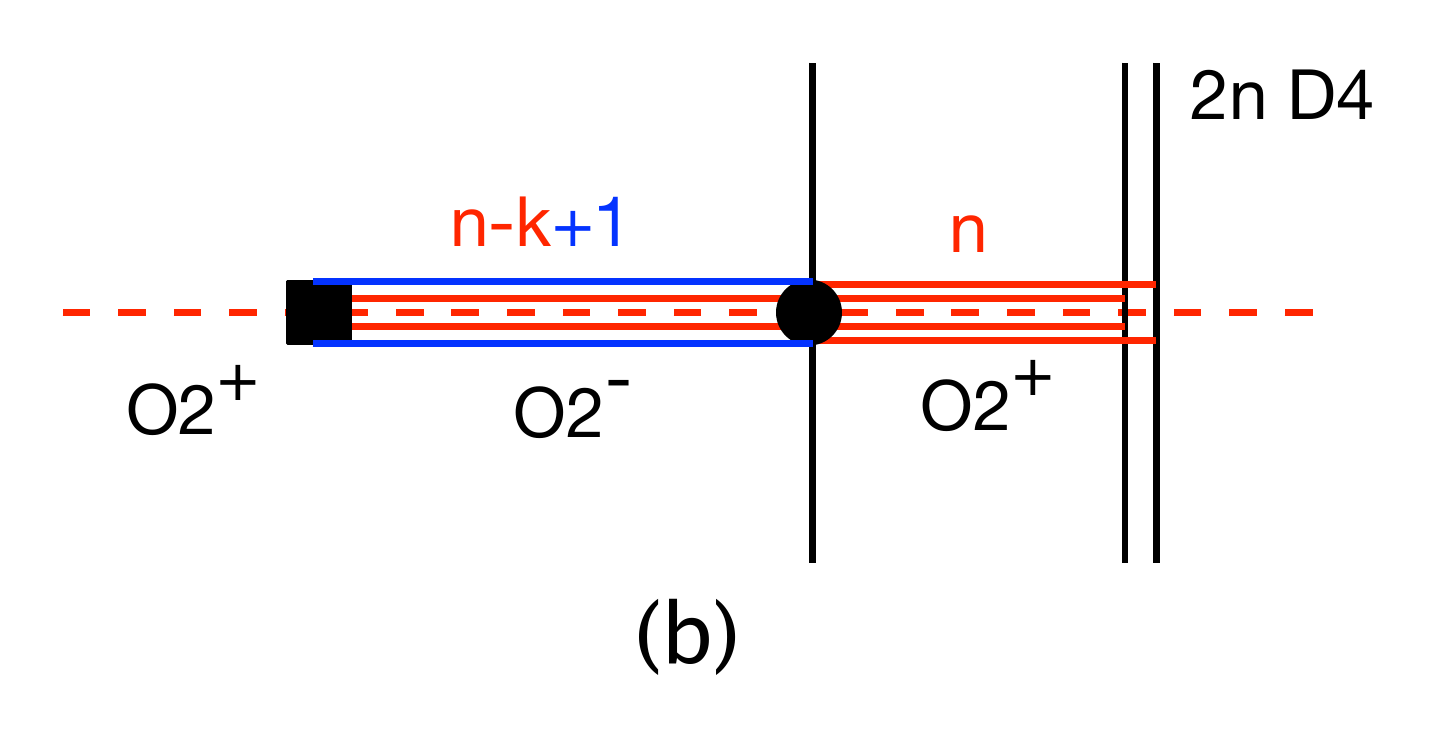}
\caption{Duality move in the Type IIA brane configuration for $O(2k)_0 + 2n+1$.}
\label{Orientifold4}
\end{figure}

Now we can revisit the exchange puzzle in the Type IIA setup.
The point is that in the Type IIA configurations of the {\em regular} theories the NS5'-brane is replaced by an NS5'-D4 bound state.
This restores the conservation of the linking number under the exchange of the two NS5-branes.
For example in the configuration for $O(2k)_\pi +2n$, Fig.~\ref{Orientifold3}, the linking number before the exchange is given by
\be 
L_{NS5} = \frac{1}{2}(N^R_{D4} - N^L_{D4}) + N^L_{D2} - N^R_{D2} =  \frac{n}{2} + \left( k -\frac{1}{8}\right) - \left( n + \frac{1}{8}\right) = k - \frac{n}{2} - \frac{1}{4} \,,
\ee
and the linking number after the exchange is given by
\be
L'_{NS5} =  \frac{n}{2} + \frac{1}{8}- \left(n-k + \frac{3}{8}\right) = k - \frac{n}{2} - \frac{1}{4} \,.
\ee
In particular no D2-branes are created in this case.
On the other hand for $O(2k)_0 + 2n+1$, Fig.~\ref{Orientifold4}, the extra D4-brane attached to the NS5'-D4 bound state leads to the creation 
of a D2-brane, and the two linking numbers are given by
\be 
L_{NS5} &=&  \frac{1}{2}\left(n - \frac{1}{2}\right) +\left(k - \frac{1}{8}\right)-\left(n + \frac{1}{8}\right) =  k - \frac{n}{2} - \frac{1}{2} \nonumber \\
L'_{NS5} & =& \frac{1}{2}\left(n + \frac{1}{2}\right) + \frac{1}{8} - \left(n-k+1- \frac{1}{8}\right) = k - \frac{n}{2} - \frac{1}{2} \,.
\ee
It is straightforward to generalize this to all the other cases in (\ref{USpDuality})-(\ref{ODualityOddOdd}).

\section{${\cal N}=(4,4)$ theories}

Much of the above analysis carries over to ${\cal N}=(4,4)$ theories in a straightforward way,
by replacing the NS5'-brane (or M5'-brane) with another NS5-brane (or M5-brane).
However the conclusions appear to be new, and somewhat unusual from the point of view 
of higher dimensional theories with eight supersymmetries.

The ${\cal N}=(4,4)$ theories include in addition an adjoint chiral superfield $X$, 
which combines with the ${\cal N}=(2,2)$ vector multiplet into an ${\cal N}=(4,4)$ vector multiplet.
The matter chiral superfields come in pairs $\Phi, \tilde{\Phi}$ transforming in conjugate representations of
the gauge group, and there is a superpotential of the general form $W \sim \Phi X \tilde{\Phi}$.
The R symmetry is $SO(4)\times SU(2)_R$.
For the unitary theory there is also a linear superpotential $W = s\,\mbox{Tr}X$ that combines with the 
twisted superpotential (\ref{FItwistedsuperpotential}) to give an $SU(2)_R$ triplet of FI terms.

The Coulomb branch is parameterized by the scalars in the vector multiplet $(\sigma, \phi_X)$, which transform
as $({\bf 2},{\bf 2},{\bf 1})$, and the Higgs branch is parameterized by
the hypermutiplet scalars $(\phi_i,\tilde{\phi}_i)$, which transform as $({\bf 1},{\bf 1},{\bf 2}) \oplus ({\bf 1},{\bf 1},{\bf 2})$.
Since the YM coupling may be regarded as the VEV of a scalar in a background vector superfield,
it cannot affect the Higgs branch, and therefore the Higgs branch does not receive quantum corrections.
On the other hand the metric on the Coulomb branch does get corrected at one-loop, as shown in \cite{Diaconescu:1997gu}.
In the IR the two branches decouple, and the theory flows to distinct superconformal theories with an $SU(2)\times SU(2)$ 
R-symmetry \cite{Witten:1997yu}.

For $G=U(k)$ with $N$ chiral superfields in the fundamental and $N$ in the antifundamental the 
${\cal N}=(4,4)$ superpotential is given by
\be
W_{U(k)} = s\, \mbox{Tr} X + \Phi_i^a X_{ab} \tilde{\Phi}_i^b \,,
\ee
and the effective twisted superpotential is given by
\be 
\widetilde{W}_{U(k)} = (t + i N\pi)\sum_{a=1}^k \Sigma_a\,.
\ee

For $G=O(K)$ with $N$ chiral superfields in the (real) vector representation the superpotential is
\be
W_{O(K)} = \Phi_i^a X_{ab} {\Phi}_j^b J^{ij} \,.
\ee
This breaks the $U(N)$ global symmetry of the ${\cal N}=(2,2)$ theory to $Sp(N)$,
and in particular requires $N$ to be even, $N=2n$.
The effective twisted superpotential is given by 
\be
\widetilde{W}_{O(K)} = i(\theta_D + \pi(2n+2K)) \sum_{a=1}^k \Sigma_a 
= i\theta_D \sum_{a=1}^k \Sigma_a \,.
\ee
This shows that at least the part of the Coulomb branch parameterized by $\Sigma$ is lifted for $\theta_D = \pi$.
Given the ${\cal N}=(4,4)$ supersymmetry we expect $X$ to be similarly lifted by an effective linear superpotential,
although we do not how to realize this in the field theory given the usual non-renormalization theorem.

For $G=USp(2k)$ with $N$ chiral superfields in the (pseudoreal) fundamental representation the superpotential is
\be
W_{USp(2k)} = \Phi_i^a X_{ab} {\Phi}_j^b \delta^{ij} \,.
\ee
This breaks the global $U(N)$ symmetry to $O(N)$.
In particular $N$ can be even or odd, corresponding to an even or odd number of half-hypermultiplets.
The effective twisted superpotential is given by 
\be
\widetilde{W}_{USp(2k)} = i\pi N \sum_{a=1}^k \Sigma_a \,,
\ee
which implies that the Coulomb branch is lifted if $N$ is odd.
Again, there should also be an effective linear superpotential for $X$, but we do not understand how it is generated.
The lifting of the Coulomb branch in this case and in the orthogonal theories with $\theta_D=\pi$
sounds a bit surprising, given the fact that in four dimensional theories with 8 supersymmetries
the Coulomb branch cannot be lifted.

The Type IIA/M theory brane construction of these theories is 
essentially identical to the one for the ${\cal N}=(2,2)$ theories, and was originally given for the $U(k)$ theory by John Brodie in \cite{Brodie:1997wn}.
We simply replace the NS5'-brane (or M5'-brane) by another NS5-brane (or M5-brane).
In particular this construction supports our assertion that the entire Coulomb branch is lifted in the cases discussed above,
since, as in the ${\cal N}=(2,2)$ construction, the M2-branes break on the flavor $\widetilde{\mbox{M5}}$-branes and cannot move
in the $(2,3,4,5)$ directions.

This construction also suggests the following dualities for the ${\cal N}=(4,4)$ theories,
\be
\label{UnitaryDuality}
U(k) + N &\longleftrightarrow & U(N-k) + N \\
\label{OrthogonalDuality}
O(K)_\pi + n &\longleftrightarrow & O(2n-K+1)_\pi + n \\
\label{SymplecticDuality}
USp(2k) + n+\frac{1}{2} &\longleftrightarrow & USp(2n-2k) + n+\frac{1}{2} \,,
\ee
where the number of flavors counts full hypermultiplets. 
The duality for $U(k)$ was originally proposed in \cite{Brodie:1997wn}, 
and tested by comparing the elliptic genera in \cite{Benini:2013xpa}.
The orthogonal and symplectic dualities are new.
There are no singlets in the magnetic theories in these cases. 

More specifically the dualities in (\ref{UnitaryDuality})-(\ref{SymplecticDuality}) should be understood as
Higgs branch dualities, namely that the two ${\cal N}=(4,4)$ supersymmetric gauge theories flow to the same IR SCFT on the Higgs branch.
Naively this does not seem possible since the condition for the existence of a pure Higgs branch for the electric theory is 
incompatible in general with the condition for the magnetic theory. 
For example for the unitary theories in (\ref{UnitaryDuality}) the electric theory requires $N\geq 2k$ whereas the magnetic theory 
requires $N\leq 2k$. One of the two theories will only have a mixed branch, in which a subgroup of the gauge symmetry remains unbroken.
However by turning on the FI parameters we can break the gauge symmetry on both sides completely for $N\geq k$,
lifting the directions in the mixed branch corresponding to nonzero VEVs for the vector multiplet scalars, thereby reducing it to a pure Higgs branch.
The dimensions of the Higgs branches, or equivalently the Higgs branch central charges \cite{Witten:1997yu}, 
of the unitary electric and magnetic theories are the same \cite{Brodie:1997wn}:
\be
\hat{c}_m = (N-k)N - (N-k)^2 = kN - k^2 = \hat{c}_e \,.
\ee
There are no FI parameters in the orthogonal and symplectic theories of (\ref{OrthogonalDuality}) and 
(\ref{SymplecticDuality}), but, as we argued above, the vector multiplet scalars are lifted in these theories.
The resulting pure Higgs branches are the same for the dual pairs.
For the orthogonal theories in (\ref{OrthogonalDuality}):
\be
\hat{c}_m = (2n-K+1)n - \frac{1}{2}(2n-K+1)(2n-K)  = Kn - \frac{1}{2}K(K-1) = \hat{c}_e \,,
\ee
and for the symplectic theories in (\ref{SymplecticDuality}):
\be
\hat{c}_m = (2n-2k)(n+\frac{1}{2}) - \frac{1}{2}(2n-2k)(2n-2k+1)  = 2k(n + \frac{1}{2})
 - \frac{1}{2} 2k(2k+1) = \hat{c}_e .
\ee

\subsection{Type IIB construction}

The ${\cal N}=(4,4)$ theories can also be realized in Type IIB string theory using D1-branes and D5-branes.
For $k$ D1-branes and $N$ D5-branes this gives a $U(k)$ theory with $N$ fundamental hypermultiplets.
The theta parameter in this description corresponds to the Type IIB RR 0-form potential, $C_0 = \theta$.
This is seen in the coupling $C_0 \mbox{Tr}F$ in the D1-brane worldvolume theory.\footnote{It is not known
how to describe the triplet FI parameter in this construction.}

The orthogonal and symplectic theories are obtained by adding an orientifold 5-plane.
There are four possibilities, $\mbox{O5}^-$, $\mbox{O5}^+$, $\widetilde{\mbox{O5}}^-$ and $\widetilde{\mbox{O5}}^+$,
that lead to different 2d gauge theories.
With the $\mbox{O5}^-$-plane the gauge group is $USp(2k)$, and there are an even number $N=2n$ of 
half-hypermultiplets in the fundamental representation.
With the $\widetilde{\mbox{O5}}^-$-plane the 2d gauge group is the same, but there is an odd number
$N=2n+1$ of half-hypermultiplets, since there is an unpaired D5-brane stuck in the orientifold plane.
The $\mbox{O5}^+$ and $\widetilde{\mbox{O5}}^+$ planes both give a 2d $O(K)$ gauge theory 
with an even number of half-hypermultiplets $N=2n$.
In this case $K$ may be even or odd, the latter corresponding to an unpaired D1-brane.
The two theories differ in the value of the discrete theta parameter:
$\theta_D = 0$ for $\mbox{O5}^+$ and $\theta_D = \pi$ for 
$\widetilde{\mbox{O5}}^+$.\footnote{The 2d ${\cal N}=(4,4)$ theories realized in this way actually contain an 
additional hypermultiplet, transforming in the 
symmetric tensor representation in
the $O(K)$ theory, and in the antisymmetric tensor representation in the $USp(2k)$ theory.
Therefore they are not identical to the theories we considered above. However the additional field does not
contribute to $\widetilde{W}$, and therefore does not change the conclusions about the Coulomb branch.}

The different versions of the orientifold 5-plane
are associated to torsion classes of RR and NSNS flux in the reduced space \cite{Hanany:2000fq}.
The $\mbox{O5}^-$-plane corresponds to trivial fluxes.
The $\mbox{O5}^+$-plane corresponds to NSNS flux in the non-trivial class of 
$H^3(RP^3,\widetilde{\mathbb Z}) = \mathbb{Z}_2$,
the $\widetilde{\mbox{O5}}^-$-plane corresponds to RR flux in the non-trivial class of
$H^1(RP^3,\widetilde{\mathbb Z}) = \mathbb{Z}_2$, and 
the $\widetilde{\mbox{O5}}^+$-plane corresponds to both fluxes being turned on.
The RR flux, in particular, corresponds to a discrete remnant of the Type IIB RR scalar potential $C_0$.
This field is odd under worldsheet parity reversal and so should reverse sign across the orientifold plane.
However it is also periodic $C_0 \sim C_0 + 2\pi$, so there are two allowed values with a vanishing field strength, 
$C_0 = 0$ or $C_0 = \pi$. 
For $\mbox{O5}^\pm$  $C_0=0$, and for $\widetilde{\mbox{O5}}^\pm$  $C_0=\pi$.
Thus in the orthogonal theory $C_0$ corresponds to the discrete theta parameter, and in the $USp(2k)$ theory
to whether the number of half-hypermultiplets is even or odd (see Table~\ref{O5planes}).

\begin{table}[h!]
\begin{center}
\begin{tabular}{|l|l|l|l|}
 \hline 
 O5-plane & $\int B_2$ & $C_0$ & 2d gauge theory \\
 \hline 
 $\mbox{O5}^-$ & 0 & 0 & $USp(2k) + n$\\
 $\mbox{O5}^+$ & $\pi$ & 0 & $O(K)_0 +n$\\
 $\widetilde{\mbox{O5}}^-$ & 0 & $\pi$ & $USp(2k) + n + \frac{1}{2}$\\
 $\widetilde{\mbox{O5}}^+$ & $\pi$ & $\pi$ & $O(K)_\pi + n$\\ 
 \hline
 \end{tabular}
 \end{center}
\caption{The four versions of the orientifold 5-plane and the corresponding D1-brane theories.}
\label{O5planes}
\end{table}

The Coulomb branch of the 2d gauge theory corresponds to moving the D1-branes away from the orientifold 5-plane.
This breaks the D1-brane gauge symmetry to $U(k)$, or more generally to $U(1)^k$, and couples the gauge field to
the RR scalar via $C_0 \mbox{Tr} F$.
Thus a non-trivial theta parameter $\theta_{eff}=\pi$ is generated on the Coulomb branch 
in the $O(K)$ theory with a non-trivial discrete theta parameter, and in the $USp(2k)$ theory
with an odd number of half-hypermultiplets, in agreement with the field theory results.
This is actually a novel phenomenon from the point of view of D-brane dynamics.
The non-trivial effective theta parameter implies that the D1-branes cannot be separated supersymmetrically
from the $\widetilde{\mbox{O5}}^-$ and $\widetilde{\mbox{O5}}^+$ planes, although naively the objects appear
to preserve eight supersymmetries independently of the relative positions.
The separation generates an electric field $F_{01} \sim \theta^2$, and therefore a non-vanishing energy density.
This again supports our assertion that the entire Coulomb branch is lifted in these cases, including the moduli
contained in the adjoint chiral superfield $X$.

\section{Discussion}

We have provided a brane realization of two-dimensional ${\cal N}=(2,2)$ supersymmetric gauge theories with
orthogonal and symplectic gauge groups, and used it to exhibit the prominent IR properties of the theories,
including Seiberg dualities.
The string theory brane configuration presented a number of puzzles, which were resolved by lifting the configuration to M theory.
This also led to some new results related to brane dynamics in M theory in the OM2-plane background
and its reduction to Type IIA string theory.

We have also discussed the ${\cal N}=(4,4)$ supersymmetric version of this construction.
This leads in particular to new duality conjectures for the {\em regular} ${\cal N}=(4,4)$ theories with orthogonal and symplectic gauge groups.
It would be interesting to further test these dualities.
It is also important to understand the mechanism that lifts the part of the Coulomb branch corresponding to the adjoint chiral superfield $X$
in these theories.

The question of the precise identification of the orthogonal gauge theory in the brane construction remains open.
Three distinct theories were considered in \cite{Hori:2011pd}, $O(K)_\pm$ and $SO(K)$.
We do not know which version the brane configuration describes.
In fact there are other possibilities such as $Spin(K)$ and $Pin(K)$.
It would be interesting to clarify this point.

\section*{Acknowledgements}
This work is supported in part by the Israel Science Foundation under grant no. 352/13,
and by the US-Israel Binational Science Foundation under grant no. 2012-041.
O.B. also thanks the Aspen Center for Physics where this project originated.

\end{document}